# Bayesian Inference on the Isotopic Building Blocks of Mars and Earth

Nicolas Dauphas[1*], Timo Hopp[1†], David Nesvorný[2]

[1]Origins Lab, The University of Chicago, 5734 South Ellis Ave, Chicago, IL 60637, USA
[2]Department of Space Studies, Southwest Research Institute, 1050 Walnut St., Suite 300, Boulder, CO 80302, USA

[*]To whom correspondence should be addressed (dauphas@uchicago.edu)

[†]Now at Now at Max-Planck-Institut für Sonnensystemforschung, 37077 Göttingen, Germany

10987 words, 9 tables, 10 figures

**Abstract**

Elements with differing siderophile affinities provide insights into various stages of planetary accretion. The isotopic anomalies they exhibit offer a deeper understanding of the materials responsible for the formation of terrestrial planets. By analyzing new iron isotopic anomaly data from Martian meteorites and drawing insights from published data for O, Ca, Ti, Cr, Fe, Ni, Sr, Zr, Mo, Ru, and Si, we scrutinize potential changes in the isotopic composition of the material accreted by Mars and Earth during their formation. Our methodology employs a Bayesian inference method, executed using a Markov Chain Monte Carlo (MCMC) algorithm.

The Bayesian method helps us balance the task of emulating the compositions of the terrestrial planets (reflected in the likelihood) with the flexibility to explore the isotopic compositions of mixing components within permissible intervals (indicated by the priors of the nuisance parameters). A Principal Component Analysis of isotopic anomalies in meteorites identifies three main clusters (forming the three parts of the isotopic trichotomy): CI, CC=CM+CO+CV+CR, and NC=EH+EL+H+L+LL. We adopt CI, COCV, O, and E as endmember compositions in the mixtures. We are concerned here with explaining isotopic anomalies in Mars and Earth, not their chemical compositions. Previous studies have shown that Earth's chemical composition could not be well explained by solely considering undifferentiated meteorites, as it requires incorporation of a refractory component enriched in forsterite. The endmember chondrite components considered here are assumed to be representative of isotopic reservoirs that were present in the solar nebula, but the actual building blocks could have had different chemical compositions, and we use cursive letters to denote those putative building blocks ($\mathcal{CI}$, $\mathcal{COCV}$, $\mathcal{O}$, and $\mathcal{E}$).

Because Earth's mantle composition is an endmember for some isotopic systems, it cannot be reproduced exactly by considering known chondrite groups only and requires involvement of a component that is missing from meteorite collections but is likely close to enstatite meteorites. With this caveat in mind, our results suggest that Earth is primarily an isotopic mixture of ~92% $\mathcal{E}$, 6 % $\mathcal{CI}$, and <2% $\mathcal{COCV}$ and $\mathcal{O}$. Mars, on the other hand, appears to be a mixture of ~65% $\mathcal{E}$, 33% $\mathcal{O}$, and <2% $\mathcal{CI}$ and $\mathcal{COCV}$. We establish that Earth's $\mathcal{CI}$ contribution substantially increased during the latter half of its accretion. Mars began accreting a mix of $\mathcal{O}$ and $\mathcal{E}$ but predominantly accreted $\mathcal{E}$ later.

Mars' changing isotopic makeup during accretion can be explained if it underwent gas-driven type I migration from its origin near the $\mathcal{O}$ - $\mathcal{E}$ boundary to a location well within the $\mathcal{E}$ region during the first few million years of solar system history. Earth's later increased $\mathcal{CI}$ contribution may be attributed to the stochastic impact of an interloper carbonaceous embryo that moved inside the inner solar system region while nebular gas was still present, and subsequently participated in the stage of chaotic growth.

The discovery that a significant portion of Earth's building blocks closely resembles enstatite chondrites contrasts with recent findings of Si isotopic anomalies in enstatite chondrites when compared to terrestrial rocks. We suggest this discrepancy likely stems from insufficient correction for high-temperature equilibrium isotopic fractionation, whether of nebular or planetary origin. With appropriate adjustments for this influence, both the silicate Earth and enstatite chondrites exhibit comparable Si isotopic anomalies, reaffirming a genetic link between them.

## 1. Introduction

Initial studies connecting terrestrial planets with potential meteoritic building blocks focused on chemical comparisons (Allègre, Manhès and Lewin, 2001; Allegre et al., 1995; Hart and Zindler, 1986; Jagoutz et al., 1979; Wänke, 1981). While it is chemically easier to build Earth from material akin to carbonaceous chondrites (Allègre, Manhès and Lewin, 2001), Javoy (1995; 1997) recognized that enstatite chondrites were the only meteorites that matched the nitrogen and oxygen isotopic compositions of terrestrial mantle rocks and postulated a connection between the two. A difficulty with this scenario is that enstatite chondrites have very low Mg/Si atom ratios (~0.75 for EH and ~0.85 for EL, Lodders, 2021) compared to Earth's mantle (~1.25, McDonough, 2003; ~1.21, O'Neill, 2014). A pure enstatite chondrite model of Earth requires a mantle Mg/Si ratio of 1.09 (Javoy et al., 2010a). This is much lower than current bulk silicate Earth (BSE) estimates (McDonough, 2003; O'Neill, 2014) and would require the existence of a volumetrically significant (~2/3 of the mantle) Si-rich (Mg/Si~1.02) reservoir in the deep mantle that has so far evaded detection (Javoy et al., 2010a). Earth's mantle also has silicon isotopic composition ($\delta^{30}$Si) that is clearly distinct from enstatite chondrites (Fitoussi and Bourdon, 2012).

Numerous elements have been shown to carry isotopic anomalies at a bulk scale (Clayton, 2003; Clayton and Mayeda, 1996; Dauphas, Marty and Reisberg, 2002a; Dauphas and Schauble, 2016b; Qin and Carlson, 2016; Trinquier, Birck and Allegre, 2007; Warren, 2011a). Isotopic analyses of a variety of elements have demonstrated that Earth's isotopic composition is close to that of enstatite chondrites, ascertaining the genetic relationship between the two (Dauphas, 2017). Some differences exist between the isotopic compositions of terrestrial rocks and enstatite chondrites for some elements for which Earth is an endmember (Fe, Zr, and Mo) but these differences are small (Budde, Burkhardt and Kleine, 2019; Render et al., 2022; Schiller, Bizzarro and Siebert, 2020). One way to reconcile the isotopic similarity between Earth and enstatite chondrites with their distinct Mg/Si ratios and Si isotopic compositions is to argue that these objects formed in the same reservoir, but that physical separation between forsterite and gas in the solar nebula produced high Mg/Si-$\delta^{30}$Si dust that through agglomeration made planetesimals precursor to the Earth, while the low Mg/Si- $\delta^{30}$Si gas reservoir produced planetesimals akin to E-type asteroids (Dauphas et al., 2015; Morbidelli et al., 2020). The parent-body of angrites has high $\delta^{30}$Si (Dauphas et al., 2015; Pringle et al., 2014) and Mg/Si (Tissot et al., 2022) presumably resulting from physical separation between forsterite and gas, and the same mechanism could have affected Earth's building blocks (Dauphas, 2017; Dauphas et al., 2015; Morbidelli et al., 2020). Contrasting with the hypothesis of a link between Earth and enstatite chondrites, an alternative view proposes that mixtures of differentiated meteorites and chondrites such as SNC and CV (Fitoussi, Bourdon and Wang, 2016) or non-aubrite achondrites (NAA) and CI (Onyett et al., 2023) can emulate Earth's isotopic composition for many elements.

Dauphas (2017) introduced a novel approach to study the isotopic nature of material accreted by Earth during its formation by leveraging the fact that elements with varying siderophile behaviors were delivered to the mantle at different stages of Earth's accretion. The conclusion of that study is that the isotopic nature of material accreted by Earth remained relatively constant, an observation more in line with a single dominant component rather than a diverse blend of components coming from different heliocentric distances. However, that study had limitations, chiefly due to the limited availability of isotopic anomaly measurements. The wealth of data produced since then enables us to reevaluate the nature of Earth's accreted material's with enhanced precision and deeper insights.

The accretion history of Mars can also be elucidated using isotopic anomalies (Brasser, Dauphas and Mojzsis, 2018; Burkhardt et al., 2021; Kleine et al., 2023; Lodders and Fegley, 1997; Paquet, Sossi and Moynier, 2023; Sanloup, Jambon and Gillet, 1999; Tang and Dauphas, 2014). Earlier attempts were hindered by insufficient data or imprecision, but the accumulation of new data over recent years provides a robust basis to revisit this question. Among the elements displaying isotopic anomalies, siderophile elements are particularly valuable for constraining the later stages of accretion. Considering this, we have measured the Fe isotopic compositions of several Martian meteorites, providing a more precise definition of the isotopic nature of Mars' accreted material.

Past studies have shown that no perfect alignment exists between the isotopic compositions of Earth or Mars and any meteorite mixture. In particular, the terrestrial mantle has an endmember isotopic composition for some elements, which cannot be perfectly reproduced by considering existing meteorites. Bayesian inference methodologies allow us to relax the strict assumption that meteorites represent an exhaustive sampling of terrestrial building blocks. Indeed, they provide an avenue to balance two crucial considerations: (*i*) making sense of observations using a model, and (*ii*) accepting that our understanding of some parameters is far from perfect and subject to change. This study uses such an approach based on the Metropolis-Hastings Markov-Chain Monte Carlo (MCMC) procedure (Gelfand and Smith, 1990; Hastings, 1970; Metropolis et al., 1953). This statistical methodology allows us to draw insightful comparisons between the accretion histories of Mars and Earth within the context of planetary dynamics.

## 2. Samples and methods

*2.1. Martian samples, chemical purification, and iron isotopic analyses*

In this study four Martian meteorites from the three major groups (shergottites, nakhlites, chassignites) were selected for Fe isotopic analyses. All four analyzed samples are meteorite falls that were previously investigated for their Ni isotopic compositions (Tang and Dauphas, 2014). Iron was purified from digestion aliquots produced in the previous study using the separation scheme described in Tang and Dauphas (2012a) that efficiently separates Fe from Cu, Ni, Co, and Cr by loading sample aliquots in 0.25 mL of 10 M HCl onto 3 mL pre-cleaned AG1-X8 (200-400 mesh) anion resin filled into 10.5 cm long PFA columns (0.62 cm ID). Iron is fixed on the resin, while Ni and other major elements are eluted in 5 mL of 10 M HCl. Other possible contaminants like Cu and Cr are rinsed from the resin using 30 mL of 4 M HCl. Finally, Fe is eluted using 9 mL of 0.4 M HCl, dried down, re-dissolved in 10 M HCl, and the chemical purification is repeated using new resin. This procedure produces purified Fe solutions that are suitable for high precision-high accuracy analyses of Fe isotopic anomalies in extraterrestrial materials (Hopp et al., 2022a; Hopp et al., 2022b).

Iron isotopic measurements were performed at the University of Chicago using a Thermo Scientific Neptune multicollector inductively coupled plasma mass spectrometer (MC-ICP-MS) that was upgraded to Neptune Plus specifications. The setup of the measurements is described in detail in Hopp *et al*. (2022a,b). The Measurements were made on flat-topped peak shoulders in medium resolution to resolve molecular argide interferences. All Fe isotopes, as well as and $^{53}$Cr and $^{60}$Ni used to monitor isobaric interferences, were measured statically in multi-collection on Faraday cups attached to $10^{10}$ Ω ($^{56}$Fe), $10^{12}$ Ω ($^{54}$Cr and $^{58}$Ni), and $10^{11}$ Ω (all other isotopes) amplifier resistors. Each measurement consisted of 50×8.369 s cycles. Solutions of 10 µg/g Fe in 0.3 M $HNO_3$ were introduced into the MC-ICP-MS using an ESI PFA nebulizer through a cyclonic glass spray chamber (Stable Introduction System) at an uptake rate of ~100 µl/min. Sample analyses were bracketed by measurements of the reference material IRMM-524a and concentrations were matched to within ≤2 %. On peak zero intensities from a blank solution measured at the beginning of each sequence were subtracted from all individual measurements and a washout time of 210 s was used between successive measurements. Possible natural mass-dependent Fe isotopic fractionation in $\delta$-notation is calculated using sample-standard bracketing,

$$\delta^{56}\text{Fe (‰)} = \left[\left({}^{56}\text{Fe}/{}^{54}\text{Fe}\right)_{\text{sample}}/\left({}^{56}\text{Fe}/{}^{54}\text{Fe}\right)_{\text{IRMM-524a}} - 1\right] \times 10^3. \qquad (1)$$

Iron isotopic anomalies were calculated by correcting for instrumental and natural mass fractionation using the exponential law (Dauphas and Schauble, 2016b; Maréchal, Télouk and Albarède, 1999) and internal normalization to $^{57}$Fe/$^{56}$Fe=0.023095, the certified ratios of IRMM-014. Isotopic anomalies are reported in the µ-notation,

$$\mu^i\text{Fe} = \left[\left({}^{i}\text{Fe}/{}^{56}\text{Fe}\right)_{(57/56),\text{sample}}/\left({}^{i}\text{Fe}/{}^{56}\text{Fe}\right)_{(57/56),\text{IRMM-524a}} - 1\right] \times 10^6, \qquad (2)$$

where the subscript (57/56) indicates the ratio used for internal normalization. The purified solutions are measured 12 to 15 times in a sample-standard bracketing scheme and the average µ values are calculated. Uncertainties for individual samples are based on repeat measurements of each sample solution using the Student's *t*-value for a two-sided 95% confidence interval and the relevant number of degrees of freedom.

2.2. *Isotopic mass balances of Mars and Earth in multi-stage accretion models*

For non-volatile elements, evaluations of the building blocks of Earth or Mars often consider mass-balance without any regard for the siderophile character of the chemical elements (Dauphas et al., 2014a; Fitoussi, Bourdon and Wang, 2016; Lodders and Fegley, 1997; Sanloup, Jambon and Gillet, 1999; Tang and Dauphas, 2014). Simulations of terrestrial planet accretion however indicate that the nature of the material accreted by Earth and Mars could have changed through time (Brasser, Dauphas and Mojzsis, 2018; Morbidelli et al., 2000; Rubie et al., 2015b). Dauphas (2017) devised an approach that uses elements with different siderophile characters to probe the isotopic composition of the material accreted by Mars and Earth through time. While lithophile elements record the full accretion history of the planet, siderophile elements provide a record that is skewed towards later accretion stages because elements delivered early were scavenged into the core (Figs. 1, 2). For example, highly siderophile elements only record the last ~0.5 of Earth's and ~0.8% of Mars's accretion histories (Day, Brandon and Walker, 2016). Under the assumptions of small impactors, constant metal-silicate partitioning, and constant metal/silicate proportions, the fraction of an element in the present mantle delivered when the planet grew from mass fractions $x_1$ to $x_2$ relative to the final size of the planet is (Dauphas, 2017),

$$X_{1\to 2} = x_2^{\kappa+1} - x_1^{\kappa+1}, \quad (3)$$

with $\kappa = Dfk_{ic}/(1-f)$, where $D$ is the core-mantle (metal-silicate) distribution coefficient, $f$ is the core mass fraction, and $k_{ic}$ is the degree of equilibration of the impactor core with the target mantle. While this analytical equation makes some assumptions, Dauphas (2017) showed that it captured the behavior of elements in more complex models involving changes in degree of metal-silicate equilibration, metal fractions, and core partitioning behaviors. We calculated effective distribution coefficients $D$ based on estimated concentrations in the mantles and cores of Mars and Earth. The core mass fractions $f$ in Mars and Earth are 0.25 (Stähler et al., 2021) and 0.32, respectively. The degree of impactor core-target mantle equilibrations $k_{ic}$ are set to 1 and 0.4 for Mars and Earth, respectively. The rationale for setting $k_{ic}$ to 1 for Mars is that it is likely a stranded planetary embryo formed by the accretion of small objects (Dauphas and Pourmand, 2011b). The value of 0.4 adopted for Earth falls within the range of 0.3 to 0.8 inferred from previous studies (Nimmo, O'brien and Kleine, 2010; Rudge, Kleine and Bourdon, 2010). The values of $\kappa$ calculated for all elements considered in Mars and Earth are compiled in Tables 1 and 2.

We consider 3 mixing models (*baseline*, *4-stage*, and *achondrite*; Table 3). For the *baseline* mixing model, we follow Dauphas (2017) and Brasser *et al.* (2018) and divide Earth's and Mars' accretions into three stages: I=0-60%, II=60-99.2% (Mars) or 60-99.5% (Earth), and III=late veneer (Table 1). The choice of dividing Earth's accretion history at 60% follows from the work of Rubie et al. (2015a) who suggested that material accreted after that could have been more oxidized. For Mars, the reason for adopting the same cutoff is that Mo and Fe record the last ~40-60% of the accretion history (Fig. 1) and adopting the same cutoff as Brasser *et al.* (2018) allows us to compare our results with that study. Ideally, we would want to divide the accretion of Earth and Mars into more numerous and finer stages, but there are insufficient anomaly data on siderophile elements to justify doing so. To test the sensitivity of the Bayesian inference to stage definition, we also ran a test by dividing the accretions of Earth and Mars into *4 stages*, corresponding to: I=0-50%, II=50-75%, III=50-99.2% (Mars) or 50-99.5% (Earth), and IV=late veneer (Table 2). We also evaluate the effect of using non-aubrite *achondrites* as endmember in mixing calculations and used only 2 stage accretion models in this context: I=0-60% and II=60-100%.

The observations that we are trying to match with the mixing calculation are the isotopic anomalies (or lack thereof) of Mars' and Earth's mantles. The total number of chondrite groups considered in the mixing calculation is $n_i$. We examine $n_j$ elements affected by isotopic anomalies. Delivery of elements to the mantles of Mars and Earth occurs over $n_k$ stages. Indices $i$, $j$, and $k$ enumerate the chondrite groups, isotopic anomalies, and stages. The isotopic anomalies of the endmembers are compiled in a $n_i \times n_j$ matrix named $I$ with meteorite groups in rows and anomalies in columns (Table 4; Appendix). The corresponding uncertainties are stored in a matrix of the same dimension named $\sigma_I$ with meteorite groups in rows and 1σ anomaly uncertainties in columns (Table 4). The concentrations of each element in the meteorite group are stored in a matrix of the same dimension named $C$ with meteorite groups in rows and concentrations in columns (Table 5). The fraction of each element delivered in each stage is compiled in a $n_j \times n_k$ matrix named $X$ with elements in rows and stage in columns (Tables 1 and 2). The parameters of interest are the mass fractions of each meteorite group in each stage, which is stored in a $n_i \times n_k$ matrix named $F$ with meteorite groups in rows and stages in columns. The calculated mantle isotopic compositions of either Mars or Earth are stored in a $n_j$-long vector named $T$ (for target) and the associated uncertainty vector is noted $\sigma_T$. If we note $T[i,j]$ the $i^{th}$ row and $j^{th}$ column of matrix $T$, the isotopic composition of the mantle $T$ can be calculate by mass-balance and its uncertainty $\sigma_T$ can be calculated by recognizing that the variance of a linear combination of independent random variables is equal to the sum of the squares of the constants multiplied by the variances of the respective variables (Dauphas, 2017),

$$T[j] = \sum_{k=1}^{n_k} X[j,k]\frac{\sum_{i=1}^{n_i} F[i,k]C[i,j]I[i,j]}{\sum_{i=1}^{n_i} F[i,k]C[i,j]} = \sum_{i=1}^{n_i}\sum_{k=1}^{n_k}\frac{X[j,k]F[i,k]C[i,j]I[i,j]}{\sum_{i=1}^{n_i} F[i,k]C[i,j]}, \quad (4)$$

$$\sigma_T[j]^2 = \sum_{i=1}^{n_i}\left(\sum_{k=1}^{n_k}\frac{X[j,k]F[i,k]C[i,j]}{\sum_{i=1}^{n_i} F[i,k]C[i,j]}\right)^2 \sigma_I[i,j]^2. \quad (5)$$

*2.3. Mixing components*

We focus on chondrites as potential building blocks for Mars and Earth. The two reasons why achondrites, iron, and stony-iron meteorites are secondary in our analysis are: (*i*) isotopic anomaly data are missing due to the low abundances of some elements that were depleted by volatilization or partitioning in metal or silicate (Table 4), and (*ii*) achondrites can be affected by admixture of material with chondritic abundances of volatile and siderophile elements, so they may not be representative of the bulk body (Zhu et al., 2021). We focus our attention on meteorite groups that have been well studied, namely CI, CM, CO, CV, H, L, LL, EH, and EL. These are also the main meteorite groups considered by Wasson and Kallemeyn (1988) in their study of the chemical compositions of the major groups of chondrites. Lodders (Lodders, 2021) also included CK and CR but these are less frequently analyzed and the database is incomplete for those groups. Examination of Table 4 and previous studies (Dauphas and Schauble, 2016b; Kleine et al., 2020; Qin and Carlson, 2016; Warren, 2011b) shows that omitting those minor groups in our analysis is of little consequence because their compositions are close to those measured in the major meteorite groups. This leaves 9 meteorite groups that we can potentially consider as building blocks of Mars and Earth. We consider $n_k$ stages and $n_i$ building blocks for the accretion of the two planets. The parameters of interest that we seek to estimate are the mass fractions of building block $i$ during stage $k$, which we note $F_{i,k}$. With the additional constraint $\forall k \in \{1,2,\ldots,n_k\} \sum_{i=1}^{n_i} F_{i,k} = 1$, we have $n_k(n_i - 1)$ parameters of interest. For 3 stages used in Dauphas (2017) and Brasser et al. (2018), and 9 meteorite groups, we therefore have 24 potential parameters. High quality isotopic anomaly measurements are available for $n_j = 11$ systems for Mars ($\Delta^{17}$O, $\varepsilon^{48}$Ca, $\varepsilon^{50}$Ti, $\varepsilon^{54}$Cr, $\varepsilon^{54}$Fe, $\varepsilon^{64}$Ni, $\mu^{66}$Zn, $\mu^{84}$Sr $\varepsilon^{96}$Zr, $\varepsilon^{92}$Mo, $\Delta^{95}$Mo) or $n_j = 12$ for Earth (like Mars but with $\varepsilon^{100}$Ru added; terrestrial rocks are used for normalization of isotopic anomaly measurements of extraterrestrial samples). Many of these anomalies are correlated. In our Bayesian analysis, we seek to maintain a balance between model complexity and parsimony to avoid overfitting. With 11-12 observables in our data, we find it reasonable to consider a model with a comparable number of parameters. This can be achieved by reducing the number of chondrite components. Dauphas (2017) only considered CI, COCV, O=H+L+LL, and E=EH+EL chondrites. Furthermore, he fixed the proportions of CI:COCV chondrites to a single value in the three stages, so the calculation only had 7 parameters of interest.

To rigorously evaluate potential dimensionality reduction within our problem, we applied Principal Component Analysis (PCA) to the isotopic anomaly data set. PCA is a method that transforms a dataset with possibly correlated variables into a new set of uncorrelated variables, called principal components. These are ordered by the amount of explained variance in the original dataset, enabling a reduction in dimensionality. We included meteorite groups CI, CM, CO, CV, CR, H, L, LL, EH, and EL in our analysis, using available data for $\Delta^{17}$O, $\varepsilon^{48}$Ca, $\varepsilon^{50}$Ti, $\varepsilon^{54}$Cr, $\varepsilon^{54}$Fe, $\varepsilon^{64}$Ni, $\mu^{66}$Zn, $\mu^{84}$Sr $\varepsilon^{96}$Zr, $\varepsilon^{92}$Mo, $\Delta^{95}$Mo, and $\varepsilon^{100}$Ru anomalies. In the context of a PCA designed to reduce dimensionality among meteorite groups, the meteorite groups function as variables while the isotopic anomalies are considered observations. Given the disparate scales on which isotopic anomalies are measured, we started our analysis with a Z-score normalization for each isotopic system across the different meteorite groups. Figure 3A presents the PCA produced with the *R* statistical software. We focus on the loadings, which are the coefficients assigned to each original variable in the formation of the first two principal components. Principal component 1 clearly segregates the meteorite groups into non-carbonaceous (NC=H, L, LL, EH, and EL) and carbonaceous (CC=CO, CV, CM, and CR) clusters, as previously defined based on cross-correlation diagrams (Budde et al., 2016; Dauphas and Schauble, 2016b; Kleine et al., 2020; Kruijer et al., 2017; Warren, 2011b). Principal component 2 isolates CI as a chondrite group that should not be lumped with CC, corroborating a conclusion already reached by Hopp *et al.* (2022a) based on cross-correlation diagrams involving Fe isotopic anomalies. Previous studies have shown that the contribution of CC to Earth and Mars was minimal (Brasser, Dauphas and Mojzsis, 2018; Budde et al., 2016; Burkhardt et al., 2021; Dauphas, 2017; Dauphas and Schauble, 2016b; Javoy, 1995; Javoy et al., 2010b; Kleine et al., 2020; Kleine et al., 2023; Kruijer et al., 2017; Lodders and Fegley, 1997; Martins et al., 2023; Nie et al., 2023; Paquet et al., 2023; Paquet, Sossi and Moynier, 2023; Sanloup, Jambon and Gillet, 1999; Savage, Moynier and Boyet, 2022; Steller et al., 2022; Warren, 2011b), so we use a single meteorite cluster (COCV) as representative of CC. As NC dominate the building blocks of Earth and Mars, considering ordinary (O=H+L+LL) and enstatite (E=EH+EL) chondrites separately offers more granular insight. While there are chemical differences between H, L, and LL, or EH and EL chondrites, these effects are secondary. CI emerges as a distinct group that cannot be grouped with any other meteorite category.

PCA does not incorporate uncertainties. To assess whether these could have influenced the way we defined meteorite clusters, we calculated the weighted Euclidean distance between isotopic vectors, whereby the distance between two meteorite groups $i1$ and $i2$ is defined as $d_{i2-i1} = \sqrt{\sum_{j=1}^{n_j} \left(I_{i2,j} - I_{i2,j}\right)^2 / \left(\sigma_{i2,j}^2 + \sigma_{i1,j}^2\right)}$ where $I_{i,j}$ is the isotopic composition of the $j^{\text{th}}$ system in the $i^{\text{th}}$ meteorite group and $\sigma_{i,j}$ is the associated $1\sigma$ uncertainty. Figure 3B presents a heat-map of these weighted Euclidean distances for pairs of meteorite groups. This analysis corroborates the result of the PCA. It separates meteorite groups into NC=O+E and others, and within NC a further division between O and E is evident. Among carbonaceous chondrites, CI is uniquely distinguished as a separate subgroup.

To summarize, our PCA and weighted Euclidean distance analyses suggest that for our study of the building blocks of Mars and Earth, we should consider the following groups: CI, COCV (a proxy for CC), O, and E. These components align with those used by Dauphas (2017), but unlike that study, we do not assume constant proportions of CI and COCV at all stages. Our approach reflects the consideration that CI might originate further out in the protoplanetary disk, and their integration into the terrestrial planet-forming region may have been influenced by the growth and migration of ice giant planets (Hopp et al., 2022a). For 3 or 4 accretion stages, we therefore have only 3×(4-1)=9 or 4×(4-1)=12 parameters of interest to consider.

In this study, our focus lies in the examination of isotopic anomalies on Mars and Earth rather than an analysis of their chemical compositions. Prior research has shown that the chemical and isotopic compositions of Earth for non-volatile elements cannot be solely attributed to documented undifferentiated meteorites (Budde, Burkhardt and Kleine, 2019; Burkhardt et al., 2011a; Burkhardt et al., 2021; Dauphas et al., 2015). Instead, there is a possible need for a reservoir enriched in a refractory component (Dauphas et al., 2015; Morbidelli et al., 2020), with an isotopic signature close but distinct from enstatite chondrites (Dauphas, 2017). To denote these hypothetical building blocks, which might have had isotopic compositions akin to known chondrite groups but different chemical makeups, we use the notations $\mathcal{CI}$, $\mathcal{COCV}$, $\mathcal{O}$, and $\mathcal{E}$ (CI, COCV, O, and E denote the actual meteorite groups). In our Bayesian inference analysis, we treat the isotopic compositions of these endmembers as hypothetical. They are informed by prior knowledge from meteorite measurements, and by the necessity for them to combine in ways that could give rise to the observed compositions of Earth and Mars.

We are interested in tying Earth and Mars to asteroid-size building blocks, which can help illuminate how the isotopic architecture of the inner solar system was mapped onto fully grown planets. Fitoussi *et al.* (2016) considered a model for Earth formation that involved mixing between Mars material and carbonaceous chondrite material akin to CV meteorites. More recently, Onyett *et al.* (2023) reported $\mu^{30}$Si isotopic anomalies in meteorites and found that E-chondrites had positive $\mu^{30}$Si values relative to terrestrial rocks, while most non-aubrite achondrites (HEDs, angrites, ureilites, pallasites, mesosiderites) had negative $\mu^{30}$Si values. Based on those results, they argued that Earth was made from a mixture of ~3/4 low-$\mu^{30}$Si inner solar system material sampled by non-aubrite achondrites and ~1/4 high-$\mu^{30}$Si outer solar system material akin to CI chondrites. We have therefore evaluated isotopic mixing models for Earth and Mars that consider achondrites ($\mathcal{A}$) as an endmember, in addition to the four chondrite endmembers highlighted above, and uses $\mu^{30}$Si as an additional constrain. We take HED meteorites (most likely originating from asteroid Vesta; McCord, Adams and Johnson, 1970; McSween Jr et al., 2013) as representative of this endmember because (*i*) they have similar isotopic compositions to angrites, ureilites, mesosiderites, and main-group pallasites (Table 4) and (*ii*) important isotopic data are missing for other achondrite groups (Ca, Fe, Ni, Sr, Si for acapulcoite, Ca, Cr, Fe, Ni, Sr, Zr, Si for lodranite, Ca, Fe, Ni, Sr, Zr, Si for brachinite, Ti, Cr, Fe, Ni, Sr, Si for winonaite, Fe, Ru for angrite, Fe, Sr, Si for aubrite, Sr, Zr for ureilite, Ca, Fe, Ni, Sr for mesosiderite, and Ca, Fe, Sr for main-group pallasite). Only Mo and Ru data are missing in HEDs but these two elements could have easily been overprinted by late accretion (Zhu et al., 2021) so the bulk isotopic composition of Vesta would be difficult to constrain anyhow. The late veneers in Earth and Mars are mostly constrained from Mo and Ru, which are not use in the achondrite accretion scenario. We therefore only consider 2 stages (I: 0-60%, II: 60-100%) in the mixing calculation involving HEDs. We keep the other mixing endmembers unchanged, so we have 2×(5-1)=8 parameters of interest to consider in the 2-stage *achondrite* accretion model (Table 3). We take the chemical composition of H chondrites (Lodders, 2021) as proxy for the composition of Vesta (Toplis et al., 2013). Toplis *et al.* (2013) also considered a model with 75% H chondrites and 25% CM-like chondrites. Adopting such a composition would not change the result much because H and CM chondrites have relatively similar chemical compositions and the bulk mixture would remain close to H.

2.4. *Bayesian inference and Markov Chain Monte Carlo (MCMC) approach*

Our aim is to estimate the mass fractions of each meteorite group at each stage, referred to as matrix $F$. Together with the isotopic compositions of the endmembers (matrix $I$), these mass fractions determine the isotopic composition of the mantle, represented as vector $T$. The relationship can be formulated as $T = g(F, I)$. The overarching objective is to determine matrix $F$ along with its corresponding uncertainties by contrasting our theoretical predictions of $T$ with its observed counterpart, denoted as $\tilde{T}$. Both the measured mantle and chondrite isotopic compositions possess inherent uncertainties (Table 4). Given the extensive number of parameters that need to be constrained, coupled with the uncertainties pervading both model parameters and observations, this problem is optimally addressed through Bayesian inference using the Metropolis-Hastings MCMC approach (Chib and Greenberg, 1995; Gelfand and Smith, 1990; Hastings, 1970; Metropolis et al., 1953). This approach has been used successfully in previous geochemical studies (Brasser, Dauphas and Mojzsis, 2018; Kipp and Tissot, 2022; Liu and Liang, 2017; Ptáček, Dauphas and Greber, 2020). We chose the Metropolis-Hastings algorithm over alternative MCMC approaches like Gibbs Sampling or Hamiltonian Monte-Carlo because it enjoys familiarity in the Bayesian community and is conceptually simple.

In the context of a probabilistic model and prior distributions over parameters (the prior is a probabilistic statement that encapsulates the knowledge or beliefs about a model parameter from prior studies or existing information), including both the parameters of interest ($F$) and nuisance parameters ($I$), MCMC strives to generate samples from the posterior distribution of these parameters given observed data ($\tilde{T}$). This is accomplished by constructing a Markov chain whose distribution matches the target posterior distribution. Starting at a set of parameters (denoted state), the algorithm iteratively proposes new parameters and decides on whether to accept or reject this state based on a set of rules. This is designed such that the distribution of the Markov chain aligns with the target posterior distribution. The MCMC approach leverages two key ingredients: (*i*) the likelihood, which gauges the probability of the observed data given a set of model parameters, and (*ii*) prior distributions that embody our initial beliefs about the parameters before seeing any data. The posterior is proportional to the product of the likelihood and prior, leading to a trade-off situation. The likelihood strives to adjust the parameters to fit the data optimally, while the prior attempts to steer the parameters towards regions considered plausible a priori. If the data align with our prior beliefs, there is no conflict. If they do not, it creates a tension, with the likelihood and the prior pulling the parameters in different directions. The resulting posterior distribution resolves this tension. It represents a compromise between the data (likelihood) and our prior beliefs (prior), each weighted by the strength of its assertion.

The log-likelihood is defined as $\ell = -\chi^2/2$, with $\chi^2(F, I) = \sum_{j=1}^{n_j} \left(T[j] - \tilde{T}[j]\right)^2 / \sigma_{\tilde{T}}[j]^2$ where $F$ represents the parameters of interest, $I$ represents the nuisance parameters, $\tilde{T}[j]$ is the measured target composition, $T[j]$ is the calculated target composition (Eqs. 4 and 5), and $\sigma_{\tilde{T}}[j]$ is the uncertainty of the measured target composition. The terrestrial isotopic composition is used for normalization of isotopic anomaly measurements, so in theory for Earth $\tilde{T}[j] = 0$ and $\sigma_{\tilde{T}}[j] = 0$, meaning that the $\chi^2$ is undefined. In practice, samples and terrestrial standards can be fractionated by natural processes or during purification following different laws than the one used for normalization (usually exponential; Dauphas and Schauble, 2016b; Tang and Dauphas, 2012b; Wombacher, Rehkämper and Mezger, 2004), which can create small artifact anomalies as has been documented for Ni (Steele et al., 2011), Mg (Davis et al., 2015), Ca, Ti, (Zhang et al., 2014), and Mo (Budde et al., 2023). Subtle isotopic heterogeneities have been found for Ru in terrestrial rocks that most likely reflect incomplete mixing of Earth's building blocks (Fischer-Gödde et al., 2020). No such heterogeneity has been found yet for other elements, but it could still exist at, or below, the current state-of-the-art precision of isotopic analyses. The assumption that for silicate Earth $\tilde{T}[j] = 0$ and $\sigma_{\tilde{T}}[j] = 0$ is therefore not fully warranted and we use for the silicate Earth the most precise isotopic analyses of the most representative samples (Table 4).

Although parameters of interest and nuisance parameters serve different functions in the inferential process, they undergo analogous treatment within the MCMC algorithm. The process necessitates the generation of proposals for both these parameters. The proposal values for the nuisance parameters $I$ (isotopic compositions of the endmembers) are generated by drawing from normal distributions with a mean of 0 and standard deviations equal to the 1σ error divided by a fixed value of 5 to 15, then adjusting the preceding values by these amounts. Proposal values for the parameters of interest $F$ (mass fraction of each endmember at each stage) are created by drawing random numbers from normal distributions with a mean of 0 and standard deviations of 0.0005 to 0.006. These values are then added to the mass fractions of $\mathcal{CI}$, $\mathcal{COCV}$, and $\mathcal{O}$. If the result is less than 0 or exceeds 1, a reflective boundary is applied to ensure that the proposed values stay within the 0 to 1 range. For each stage, the mass fraction of $\mathcal{E}$ is calculated as the difference between the sum of the mass fractions $\mathcal{CI}$+$\mathcal{COCV}$+$\mathcal{O}$ and 1. This proposal construction guarantees that the mass fractions of $\mathcal{CI}$, $\mathcal{COCV}$, and $\mathcal{O}$ remain within 0 and 1, and that the mass fractions of all endmembers sum to 1, all the while preserving detailed balance (for every pair of states $i$ and $j$, the probability of transitioning from $i$ to $j$ must be the same as the probability of transitioning from $j$ to $i$), a requirement of the Metropolis-Hastings algorithm . The sole drawback with these proposals is that the mass fraction of $\mathcal{E}$ can be negative. This situation is managed in the prior function where such a proposal is assigned a log-prior probability of $-\infty$. For the nuisance parameters (the isotopic compositions of the meteorite endmembers), we apply normal distributions as priors, centered on the mean with a standard deviation equal to the 1σ uncertainty of the parameter (Table 4). The priors for the mass fractions of each mixing component at each stage are uniform between 0 and 1.

We implemented the MCMC algorithm using the Mathematica software. The main steps include:

a. Finding the values of the parameters of interest that minimize the $\chi^2$. These are used as the starting state in the chain.

b. Proposing a new state for the parameters of interest and the nuisance parameters via a random walk from the current state.

c. Estimating the likelihood of the observed data for both the current and the proposed states. Multiplying these likelihoods by the prior probabilities of these states to compute the unnormalized posterior probabilities. This is a critical step in Bayesian inference as it synthesizes prior knowledge on nuisance parameters (constraints on endmember compositions from previous meteorite analyses) and parameters of interest (the mass fractions of each component at each stage must be within [0,1]), with new observed data (the isotopic compositions of Earth's and Mars' mantles as they relate to the isotopic compositions and mass fractions of the mixing endmembers).

d. Implementing the Metropolis-Hastings criterion to accept or reject the proposed state. Calculate $\alpha$, the ratio of the posterior probabilities of the proposed and current states. If $\alpha \geq 1$ the proposal is accepted. If $\alpha < 1$, the new state is accepted with a probability of $\alpha$. This is achieved by drawing a random number between 0 and 1 and accepting the new state if that number is less than $\alpha$. If the proposed is accepted, it is appended to the chain. If it is rejected, the current state is appended to the chain.

f. Repeating from step b, until the chain accumulates enough iterations.

The iterations were conducted $10^6$ times. The initial 200,000 iterations were disregarded as a part of the burn-in period to allow the Markov chain to reach a stable state. The step sizes were fine-tuned based on preliminary trials runs to achieve a rejection rate of ~40-70% and to ensure a comprehensive exploration of the parameter space. The chains were visualized and assessed to confirm convergence. Different forms of the proposal functions for the parameters (uniform, gaussian) were evaluated, yielding similar results, albeit at different convergence rates.

## 3. Results

Unless otherwise noted, all uncertainties are reported as 95% confidence intervals.

3.1. *Iron isotopic analyses of SNC meteorites*

The Fe isotopic data of the four samples show no resolvable mass-dependent Fe isotopic fractionations relative to IRMM-524a (Table 6). The internally normalized data display resolvable Fe isotopic anomalies in $\mu^{54}$Fe but no clearly resolvable isotopic anomalies in $\mu^{58}$Fe, which agrees with previous meteorite analyses (Hopp et al., 2022a,b). The four samples that represent the three major groups of Martian meteorites (shergottites, nakhlites, chassignites) display indistinguishable $\mu^{54}$Fe and $\mu^{58}$Fe values and define weighted averages of +6±3 and -5±4, respectively (±95% c.i, *n*=4; Fig. 4, Table 6). The $\mu^{54}$Fe values agree with available data from Schiller *et al.* (2020) (weighted mean of 7±3, *n*=4). No $\mu^{58}$Fe data was previously reported. Together the data from this study and previously allow to define a weighted average $\mu^{54}$Fe for bulk silicate Mars of +7±2 (95% c.i.).

*3.2. Bayesian inference*

For each mixing model, the output of the Bayesian inference calculation is the set of 800,000 states in the Markov chain that describe the posterior distribution of the parameters of interest (the mass fractions of each meteorite group at each stage) and the nuisance parameters (the isotopic compositions of the mixing endmembers). From those, we can also calculate the predicted model compositions for Earth and Mars. These predicted compositions differ from a regular best fit model obtained by a simple $\chi^2$-minimization, as the nuisance parameters are also updated in the analysis. The posterior mass fractions of each meteorite endmember at each stage are compiled in Tables 7 and 8 for Mars and Earth, respectively. The posterior probabilities of the meteorite endmember isotopic compositions in the baseline model are compiled in Table 9. The predicted composition of the mantles of Mars and Earth are compiled in Table 4. We have considered 3- (*baseline*) and *4-stage* accretion models involving mixtures of $\mathcal{CI}$, $\mathcal{COCV}$, $\mathcal{O}$, and $\mathcal{E}$ (Table 3) to assess the influence of dividing planetary accretion in discrete accretion stages. We have also considered a 2-stage accretion model involving the same components plus achondrites ($\mathcal{A}$; achondrite model; Table 3) to evaluate the role that such a component could have played in terrestrial planet accretion. Below, we start by discussing the results on the main accretion stages and we then examine the late veneer.

Earth is dominantly made of $\mathcal{E}$ material during the main stages of its accretion (Figs. 5, 6, 7; Table 8), with contributions of $\mathcal{E}$ -material to bulk Earth of 92.5 (+2.3/-2.5)%, 91.0 (+2.7/-4.0)%, and 92.9 (+2.1/-2.3)% in the *baseline*, *4-stage*, and *achondrite* models, respectively. This agrees with previous studies that had recognized the isotopic similarity between Earth's mantle and enstatite chondrites for several elements (Dauphas et al., 2014b; Javoy, 1995; Javoy et al., 2010b). Using fewer and less precise data, Dauphas (2017) and Brasser et al. (2018) had concluded that Earth was also predominantly made of $\mathcal{E}$ material but found a ~50% $\mathcal{O}$ contribution during stage I. This is not seen in any of our inferences, and the baseline model yields an $\mathcal{O}$ contribution of only 1.0 (+3.0/-0.6)% during stage I. As discussed by Dauphas (2017) and Brasser et al. (2018), the significant $\mathcal{O}$ contribution was due in part to uncertainties in model end member definitions and a higher $\mathcal{E}$-contribution was plausible. The present analysis corroborates this inference. In the achondrite model, we see no evidence for a large achondrite contribution (Table 8).

The contributions of $\mathcal{CI} + \mathcal{COCV}$ increases slightly between early and late stages of Earth's accretion (excluding the late veneer). While this is indicated in Fig. 6, examination of the MCMC posterior distribution allows us to evaluate this more rigorously as we can examine how the fraction of NC changed across different stages in individual elements of the Markov chain. If we note $f_{\mathrm{NC}}$ the fraction of NC= $\mathcal{O}$+ $\mathcal{E}$ in a given stage, examination of the posterior distributions yields $f_{\mathrm{NC,II}}/f_{\mathrm{NC,I}} = 0.91\,^{-0.10}_{-0.08}$ in the *baseline* model and $f_{\mathrm{NC,II+III}}/f_{\mathrm{NC,I}} = 0.90 \pm 0.08$ in the *4-stage* accretion model, so the fraction of $\mathcal{E}$ decreased by ~10% in the later 40% of Earth's accretion history, and $\mathcal{CI}$ made up for much of that decline.

The nature of the late veneer on Earth is poorly constrained. It is almost entirely constrained by the isotopic composition of Ru and to a lesser extent by Mo. The late veneer is mostly $\mathcal{CI}$ in the *baseline* model and mostly $\mathcal{E}$ in the *4-stage* model (Fig. 5). Only one CI meteorite had its $\varepsilon^{100}$Ru composition measured at -0.24±0.13 (Fischer-Gödde and Kleine, 2017), while E-chondrites define an average of -0.08±0.04 (Fischer-Gödde et al., 2015; Fischer-Gödde and Kleine, 2017). $\mathcal{E}$ is therefore favored based on the likelihood for $\varepsilon^{100}$Ru, but allowing a $\mathcal{CI}$ contribution helps fit the $\Delta^{95}$Mo value of the bulk silicate Earth, as 10 % of the mantle inventory delivered by $\mathcal{CI}$ is sufficient to increase the $\Delta^{95}$Mo value of the mantle by ~+6, which is close to the value measured in Earth (Budde, Burkhardt and Kleine, 2019).

Even when $\mu^{30}$Si and an achondritic component are considered, the Bayesian inference finds a negligible fraction of achondrites and instead moves the prior $\mu^{30}$Si of $\mathcal{E}$ from +12.5±2.2 to +7.4±1.6. We also ran a $\chi^2$ minimization to evaluate the mixture that reproduces the $\mu^{30}$Si value and fits the other isotopic anomalies in the *achondrite* 2-stage model (Table 3). To do this, we arbitrarily divide the errors on the $\mu^{30}$Si values of the silicate Earth and the mixing components by 10. In that case, the optimum mixture is 33% $\mathcal{CI}$+9% $\mathcal{COCV}$ +58% $\mathcal{A}$ during stage I and 31% $\mathcal{CI}$ +69% $\mathcal{A}$ during stage II (32% $\mathcal{CI}$ +6% $\mathcal{COCV}$ +62% $\mathcal{A}$ overall). This mixture corresponds approximately to that inferred by Onyett *et al*. (2023). The calculated isotopic anomalies are $\Delta^{17}$O=-0.21±0.06, $\varepsilon^{48}$Ca=-0.10±0.16, $\varepsilon^{50}$Ti=-0.10±0.03, $\varepsilon^{54}$Cr=-0.05±0.05, $\mu^{54}$Fe=+9.4±1.9, $\mu^{84}$Sr=+13±10, $\mu^{96}$Zr=+54±7, $\mu^{30}$Si=+2.1±0.2. While this mixture reproduces the $\mu^{30}$Si value, it fails in many respects. The situation is particularly dire for $\mu^{54}$Fe and $\mu^{96}$Zr because the silicate Earth is an endmember for these isotopic systems and both carbonaceous meteorites and non-aubrite achondrites display large isotopic anomalies (Table 4).

To summarize for Earth, most of its accretion consisted of $\mathcal{E}$, which decreased slightly in the latter 40% of the accretion while the contribution of $\mathcal{CI}$ increasing during that stage. The nature of the late veneer is uncertain, and it could either have been $\mathcal{CI}$ or $\mathcal{E}$.

In *baseline* and *4-stage* accretion models, Mars is a mixture between $\mathcal{O}$ and $\mathcal{E}$ ( $32.7^{+5.3}_{-4.9}$ % $\mathcal{O}$ + $64.8^{+5.1}_{-5.4}$ % $\mathcal{E}$ + $1.8^{+2.0}_{-1.2}$ % $\mathcal{CI}$ + $0.7^{+0.7}_{-0.5}$% $\mathcal{COCV}$ and $33.1 \pm 5.4$ % $\mathcal{O}$ + $63.6 \pm 5.5$ % $\mathcal{E}$ + $1.9^{+1.8}_{-1.3}$ % $\mathcal{CI}$ + $0.6^{+0.7}_{-0.4}$% $\mathcal{COCV}$, respectively; Figs. 5, 7, and 8). This agrees with previous studies that used fewer isotopic systems and less well-constrained isotopic endmembers, which gave 30-45% O + 55-70% E (Sanloup, Jambon and Gillet, 1999) and 32-67% O + 29-68% E (Brasser, Dauphas and Mojzsis, 2018). The main novelties of the present work are that we can better define the mixing proportions and can discern changes in the proportions of $\mathcal{O}$ and $\mathcal{E}$ accreted by Mars at different stages of its growth. The mixing proportions inferred by Lodders and Fegley (1997) of 85% H+11% CV+4% CI cannot explain the isotopic composition of Mars (Fig. 2 of Tang and Dauphas, 2014).

The contribution of $\mathcal{CI}$+$\mathcal{COCV}$ does not change significantly between the first 60% and latter 40% of Mars' accretion history, although the calculation hints at a possible increase. The most striking observation is the increase in the proportion of $\mathcal{E}$ relative to $\mathcal{O}$ in later stages of Mars' accretion history (Fig. 8). If we note $f_{\mathcal{O}}$ and $f_{\mathcal{E}}$ the fractions of $\mathcal{O}$ and $\mathcal{E}$ in a given stage, examination of individual states in the Markov chain yields $\left[f_{\mathcal{E},\mathrm{II}}/\left(f_{\mathcal{E},\mathrm{II}}+f_{\mathcal{O},\mathrm{II}}\right)\right]/\left[f_{\mathcal{E},\mathrm{I}}/\left(f_{\mathcal{E},\mathrm{I}}+f_{\mathcal{O},\mathrm{I}}\right)\right] = 1.7^{+0.7}_{-0.6}$ in the baseline model, which is statistically significant. More uncertainty is present in the *4-stage* model and we have $\left[f_{\mathcal{E},\mathrm{III}}/\left(f_{\mathcal{E},\mathrm{III}}+f_{\mathcal{O},\mathrm{III}}\right)\right]/\left[f_{\mathcal{E},\mathrm{I+II}}/\left(f_{\mathcal{E},\mathrm{I+II}}+f_{\mathcal{O},\mathrm{I+II}}\right)\right] = 1.5^{+1.3}_{-0.8}$. Brasser et al. (2018) had examined this question but the data available at that time were insufficient to see this change. The nature of the late veneer is poorly constrained in all Mars models, although a dominantly carbonaceous component is favored (Figs. 5 and 7).

When achondrites are considered as an endmember in a 2-stage accretion model for Mars using $\mu^{30}$Si as constraint, the Bayesian inference calculation finds a significant contribution of this endmember ($38.6^{+8.2}_{-12.8}$% of the bulk planet) but the mixture becomes more *ad hoc* as it requires involvement of more meteorite groups to reproduce the composition of silicate Mars (Table 7), as opposed to just two components ($\mathcal{O}$+$\mathcal{E}$) when achondrites are not considered.

The *baseline* and *4-stage* mixing models of Earth and Mars both reproduce the compositions of these planets relatively well and none stands out as superior to another (Table 4). Comparison of the posterior distributions of the isotopic compositions of the endmembers can provide us with some clues on which isotopic systems are more difficult to reproduce and what a missing endmember could look like (Table 9). In the *baseline* Mars model, the isotopic compositions that change the most between prior and posterior are: $\varepsilon^{50}$Ti of $\mathcal{E}$ from -0.16±0.06 to -0.23±0.05, and $\mu^{84}$Sr of $\mathcal{E}$ from -11±13 to -20±11. The changes are relatively modest. In the *baseline* Earth model, the isotopic compositions that change the most are: $\mu^{54}$Fe of $\mathcal{E}$ from +6.4±1.1 to +3.2±0.8, $\varepsilon^{96}$Zr of $\mathcal{E}$ from +14±5 to -2±2, $\varepsilon^{92}$Mo of $\mathcal{E}$ from +0.40±0.08 to +0.32±0.07, and $\varepsilon^{100}$Ru of $\mathcal{CI}$ from -0.24±0.13 to +0.05±0.03. The changes are significant and concern primarily isotopic systems for which Earth is an endmember, with $\mathcal{E}$ a relatively close neighbor. The only way for the posterior to improve in the inferential process is for the prior of $\mathcal{E}$ for these isotopic systems to be updated. This suggests that meteorite collections likely lack an extraterrestrial component, for which E chondrites merely serve as a representative proxy. This could affect our conclusions. Similarly, the inferential process handles the tension between reproducing $\Delta^{95}$Mo and $\varepsilon^{100}$Ru, and this is resolved by having a late veneer dominantly made of $\mathcal{CI}$ but with an $\varepsilon^{100}$Ru value that is updated in the inferential process. The $\varepsilon^{92}$Mo of $\mathcal{CI}$ changes from +0.97±0.45 to +0.36±0.41, which helps explain the positive $\Delta^{95}$Mo value of Earth's mantle relative to $\mathcal{E}$ by allowing a significant late contribution of $\mathcal{CI}$. Only two specimens of the Orgueil CI chondrite have been analyzed for their Mo isotopic compositions (Burkhardt et al., 2011b; Dauphas, Marty and Reisberg, 2002b) and only one specimen of Orgueil has been analyzed for its Ru isotopic composition (Fischer-Gödde and Kleine). In both cases, the $\varepsilon^{92}$Mo and $\varepsilon^{100}$Ru values are rather imprecise, calling for more precise measurements, given the possible role of this CI chondrites in explaining the positive $\Delta^{95}$Mo value of Earth's mantle.

Recently discovered zinc (Martins et al., 2023; Paquet et al., 2023; Savage, Moynier and Boyet, 2022; Steller et al., 2022) and potassium (Nie et al., 2023) isotopic anomalies can provide complementary pieces of information on the nature of the material accreted by the Earth. Assuming chondritic composition for the building blocks, zinc isotope anomalies give a ~6 to 13 % mass of the Earth from carbonaceous outer solar system material (~30% of its Zn inventory), while the rest came from the inner solar system. Kleine et al. (2023) and Paquet *et al.* (2023) concluded that Mars' Zn inventory came primarily from the inner solar system, with CC not exceeding ~4% of Mars' mass. Potassium isotopic anomalies also indicate that in both Earth and Mars, K came predominantly from inner solar system bodies, but the data remain consistent with a carbonaceous contribution not exceeding ~10%. One challenge in interpreting isotopic anomalies of moderately volatile elements in terms of accretion history is the potential loss of these elements through volatilization, either in the accreted material itself or during planetary accretion. For instance, if a significant portion of Earth's formation derived from materials heavily depleted in moderately volatile elements, such as angrites, K and Zn isotopic records in the mantle might have missed 86 to 94 % of Earth's accretion (Earth is depleted by factors of 7.4 in K, Dauphas et al., 2022, and ~15 in Zn, McDonough and Sun, 1995). In that context, the agreement in the contribution of carbonaceous materials between refractory and moderately volatile elements is remarkable but could be a coincidence, as they could record very different stages of planetary accretion.

## 4. Discussion

The most significant conclusions drawn from existing and newly measured isotopic anomalies are the following:

- Mars accreted ~65% $\mathcal{E}$, 33% $\mathcal{O}$, and ~2% $\mathcal{CI}$+CC. In the first part of its accretion, it accreted $\mathcal{O}$ and $\mathcal{E}$ in approximately equal proportions but accreted predominantly $\mathcal{E}$ after that. The contribution of carbonaceous material may have increased in later stages, including during accretion of the late veneer, but this is highly uncertain.
- Earth accreted ~92% $\mathcal{E}$, 6% $\mathcal{CI}$, and less than 2% CC and $\mathcal{O}$. The contribution of $\mathcal{CI}$ increased from ~2% in the first part of the accretion to ~11% in the later part.

### 4.1. *Implications for planetary dynamics*

The field has experienced a significant surge in ideas related to terrestrial planet accretion (Brož et al., 2021; Chambers and Wetherill, 1998; Clement et al., 2018; Johansen et al., 2021; Levison et al., 2015; Nesvorný, Roig and Deienno, 2021; Walsh et al., 2011; Woo et al., 2023). Our primary focus here is on circumscribed phenomena that will require further examination within more comprehensive accretion models. The scenario we propose and the possible timeline of event we detail are illustrated in Fig. 9.

Radiometric dating indicates that Mars accreted rapidly, reaching half of its present size in approximately 2 Myr (Dauphas and Pourmand, 2011a; Tang and Dauphas, 2014). This observation, together with the difficulty of accounting for its small size in planetary dynamics simulations (Wetherill, 1991), indicate that Mars is most likely a stranded planetary embryo that escaped the stage of chaotic growth involving protracted collisions between Moon- to Mars-size objects (Dauphas and Pourmand, 2011a; Kobayashi and Dauphas, 2013). The protosolar nebula is thought to have lasted ~5 Myr (Dauphas and Chaussidon, 2011; Krot et al., 2005), so this would mean that Mars formed when nebular gas was still around. Processes such as gas driven type-I migration of proto-Mars, aerodynamic drag on planetesimals, and the evolution of the gas disk could have affected the nature of the material accreted by Mars at different times.

There are considerable uncertainties on how isotopic anomalies were distributed across the early solar system and the discussion below is speculative. Based on spectroscopic observations of asteroids and cosmochemical considerations, it is likely that the early solar system was characterized by different isotopic reservoirs from inside to outside of the solar system along a sequence $\mathcal{E}$, $\mathcal{O}$, $\mathcal{COCV}$ (CC), and $\mathcal{CI}$ (Brasser, Dauphas and Mojzsis, 2018; Dauphas et al., 2014b; Desch, Kalyaan and Alexander, 2018; Hopp et al., 2022a; Morbidelli et al., 2012; Render and Brennecka, 2021). We find that during the first 60% of its growth (up to ~2.2 Myr after solar system formation; Dauphas and Pourmand, 2011a), Mars accreting material was composed of an equal mixture of $\mathcal{O}$ and $\mathcal{E}$ (Fig. 5, Table 7). The feeding zone of embryos is approximately symmetrical, and a reasonable assumption is that Mars was born near the limit of the $\mathcal{E}$- and $\mathcal{O}$-planetesimal dominated regions (first panel in Fig. 9).

In the last ~40% of its accretion, Mars accreting material became predominantly $\mathcal{E}$ (Fig. 5, Table 7). This could either mean that the boundary between $\mathcal{O}$- and $\mathcal{E}$-dominated regions moved relative to the radial location of Mars. Small planetesimals could have drifted inward because of aerodynamic gas drag (Adachi, Hayashi and Nakazawa, 1976) but this would lead to an increase in the fraction of O rather than a decrease as is observed. The $\mathcal{E}$ planetesimals that formed at ~1 au could have been gravitationally scattered outward as Mars-class and larger protoplanets formed near 1 au (Brož et al., 2021). This could have plausibly increased the $\mathcal{E}/\mathcal{O}$ ratio in the Mars feeding zone over time. The scattered planetesimals could not have reached >2 au because their orbits would have been circularized at <2 au by gas drag.

Alternatively, Mars migrated radially inwards from near the $\mathcal{E}$-$\mathcal{O}$ boundary to well within the $\mathcal{E}$-region (second panel in Fig. 9). Type I migration arises from corrotation and Lindblad torques exerted on the planet from the perturbed gas disk (Goldreich and Tremaine, 1980; Ward, 1986). McNeil *et al.* (2005) studied type I migration for embryos in the terrestrial planet forming region and found that they would have to rapidly form for their orbits to avoid significant orbital decay. In their simulations, embryos found at 1.5 au would have originated at ~1.5 to 2.5 au. Brož (2021) invoked similar migration of Mars in their model of convergent migration of protoplanets toward 1 au, as may be needed to explain the tight radial spacing of Earth and Venus. It is thus conceivable that Mars started its formation closer to the main asteroid belt where it received a significant contribution of $\mathcal{O}$ but then drifted inwards on a timescale of a few million years, and thus received a larger contrition of $\mathcal{E}$ at later times.

The contribution of carbonaceous asteroids to the building of Mars remained relatively modest but could have increased in later stages, especially during the late veneer stage. Given Mars' rapid growth, the late veneer stage may have occurred relatively early, when nebular gas was still around, or within ~10 Myr of its dispersal. The growth of giant planets would have scattered carbonaceous planetesimals from ~5-20 au (CC and $\mathcal{CI}$) toward the inner solar system. The initially eccentric orbits of scattered planetesimals would have been rapidly circularized by gas drag and implanted in the main asteroid belt (Nesvorny et al., 2023; Raymond and Izidoro, 2017). Toward the end of gas disk lifetime, when the gas density at <5 au was low (Komaki et al., 2023) and the effect of gas drag was weakened, a small fraction of carbonaceous planetesimals originally from ~5-20 au could have been implanted in the Mars-feeding zone at <2 au. This could be a potential source of the carbonaceous component in Mars's late veneer. After gas disk dispersal, the asteroid belt was excited and depleted under the influence of outer planet migration/instability (Nesvorný and Morbidelli, 2012; Tsiganis et al., 2005), with a small fraction of carbonaceous and $\mathcal{O}$ asteroids impacting Mars. We do not expect this to be a major source of Mars accretion, because the asteroid belt had a relatively low mass and the impact probability of asteroids on Mars is low.

Earth's composition cannot be reproduced exactly by any chondritic mixture, but it clearly carries an $\mathcal{E}$ signature (Figs 5, 7, Table 8). There is no evidence for extensive mixing with the $\mathcal{O}$-reservoir, suggesting that it grew by accretion of local embryos. Chaotic growth of Earth would be expected to be associated with considerable mixing of embryos formed at different heliocentric distances (Chambers and Wetherill, 1998). The fact that $\mathcal{O}$ material is largely missing from Earth but represents ~50 % of Mars could indicate that the $\mathcal{O}$-region was relatively limited in extent. It was present near the birthplace of Mars, but few embryos were born with that signature. This could be understood if Mars formed near the $\mathcal{E}/\mathcal{O}$ boundary (see above), and Earth accreted planetesimals and protoplanets from ~1 au. Mixing of accretion materials from different heliocentric distances is reduced when the gas disk is still around because orbital eccentricities of planetesimals/protoplanets at ~1 au remain low.

A significant result of the Bayesian inference is that the fraction of carbonaceous material accreted by Earth likely increased later in its history (Figs 5, 6, Table 8). There is considerable debate on the timescale of Earth's accretion. Fast accretion (~10 Myr timescale) can potentially be reconciled with excess $^{182}$W from $^{182}$Hf-decay if the Moon-forming impact happened late and involved significant interaction between Theia's core and the protoEarth mantle (Yu and Jacobsen, 2011). Such a scenario would be more relevant to Earth's accretion in the protoplanetary gas disk by accretion of planetesimals or pebbles (Johansen et al., 2021; Levison et al., 2015), possibly associated with convergent migration of protoplanets (Brož et al., 2021). If Earth grew in a more canonical manner through collisions between planetary embryos, then Earth's accretion likely unfolded over a longer period, achieving 60% of its final mass in ~5 to 50 Myr (Gu et al., 2023).

In the *baseline* model, the mass of carbonaceous material accreted by Earth in the last ~40% of its accretion (excluding the late veneer) would have represented 4.5±2.5% of Earth's total mass. In the *4-stage* accretion model, the mass of carbonaceous material accreted by Earth in the last ~25% of its accretion (again excluding the late veneer) would have represented 4.7±2.4% of Earth's total mass. This finding is challenging to explain, because it is difficult to imagine how carbonaceous material could be such an important contributor during the late stages of Earth's accretion.

Planet migration and dynamical instability in the outer solar system (Tsiganis et al., 2005) is an appealing scenario to explain the late delivery of carbonaceous material to Earth, but there are significant difficulties. After gas dispersal, dynamical interaction of Neptune with trans-Neptunian disk of icy planetesimals fueled radial migration of Neptune from its presumed formation radius at ~20 au to its current orbit at 30 au. Neptune's migration could have triggered a dynamical instability in the outer solar system, known as the "Nice" model (Tsiganis et al., 2005), when Jupiter's eccentricity was excited to its current value (~0.05). This change should have destabilized the orbits of carbonaceous planetesimals that were previously implanted in the asteroid belt by gas drag (Nesvorny et al., 2023; Raymond and Izidoro, 2017). A fraction of these carbonaceous planetesimals would end up accreting on Earth. The issue with this scenario is that the terrestrial collision probability of destabilized planetesimals is relatively low, on the order of $10^{-3}$ (Gladman et al., 1997). We seek to explain delivery of ~4.5 % of Earth's total mass late in Earth's accretion, which for a ~$10^{-3}$ efficiency would require 4.5 Earth-mass worth of carbonaceous asteroids in the outer belt, which is unreasonable. The probability of direct collision with the Earth of outer solar system planetesimals flung inwards during Neptune's migration is als overy low (~$10^{-6}$; Gomes *et al.*, 2005). They should not represent a significant source for Earth's accretion either.

A more appealing scenario is that delivery of carbonaceous material to Earth took the form of one of several stochastic impacts with large embryos. While small (~100 km) and early formed carbonaceous planetesimals at ~5-10 au would be preferentially implanted in the asteroid belt (Nesvorny et al., 2023; Raymond and Izidoro, 2017), larger objects such as small embryos (~1000 km) could have reached <2 au (the effects of gas drag are reduced for larger bodies), where they would have mingled with embryos formed in the inner solar system. The implanted carbonaceous embryos would have participated in the stage of chaotic growth, and one or a few of those interlopers could have been accreted by Earth several tens of million years after solar system formation.

Earth's late veneer could either have been $\mathcal{E}$ or $\mathcal{CI}$. From a dynamic point of view, $\mathcal{E}$ would be easier to understand if it represented just the tail end of leftover accreting materials in Earth's vicinity, but the scenario outlined above could also provide some justification for a possible $\mathcal{CI}$ component of the late veneer. This is because, if $\mathcal{CI}$ planetesimals formed in the Uranus/Neptune region at >10 au (Hopp et al., 2022a) or even >15 au (Desch, Kalyaan and Alexander, 2018), and if Uranus/Neptune reached their final masses relatively late during the gas disk lifetime, the gas density at <5 au could have been already reduced (for example by MHD winds, Komaki *et al*., 2023). The $\mathcal{CI}$ planetesimals scattered by Uranus/Neptune could have been subsequently implanted to <2 au by gas drag, where they would later contribute to Mars and Earth late veneers.

The scenario proposed above can be tested against dynamical simulations to weed out those that do not reproduce the constraints that we see. Considerable uncertainties remain for the Earth because an endmember component is still missing.

### 4.2. *A significant achondritic component in Earth?*

The claim was recently made based on Si isotopic anomalies that the making of the Earth requires a significant achondritic component (Onyett et al., 2023). Enstatite chondrites were excluded by Onyett *et al.* (2023) because they display positive $\mu^{30}$Si anomalies of +11±3. Onyett *et al.* (2023) argue instead for a model that involves mixing between achondrites and CI+CC meteorites. While such a mixture can reproduce the $\mu^{30}$Si value, it fails in many respects, most notably for $\mu^{54}$Fe and $\mu^{96}$Zr because the silicate Earth is an endmember for these isotopic systems and both carbonaceous meteorites and non-aubrite achondrites display large isotopic anomalies (Fig. 10; Table 4).

Onyett et al. (Onyett et al., 2023) argue that the difference in $\mu^{30}$Si between enstatite chondrites and BSE is of nucleosynthetic origin, ruling out the possibility that the two be genetically tied by formation from the same isotopic reservoir. A potential complication however is that enstatite chondrites have very fractionated $\delta^{30}$Si values relative to BSE (Armytage et al., 2011; Dauphas et al., 2015; Fitoussi and Bourdon, 2012; Georg et al., 2007; Onyett et al., 2023; Savage and Moynier, 2013; Zambardi et al., 2013), which can create spurious isotopic $\mu^{30}$Si values if natural fractionation departs from the exponential mass fractionation law used for internal normalization (Dauphas and Schauble, 2016a; Davis et al., 2015; Tang and Dauphas, 2012a; Wombacher, Rehkämper and Mezger, 2004; Young, Galy and Nagahara, 2002). It is therefore not entirely clear if the apparent $\mu^{30}$Si anomalies measured by Onyett *et al.* (2023) are nucleosynthetic in origin or if they are artifacts arising from the use of an incorrect mass fractionation law in the data procedure. Onyett et al. (2023) dismissed this concern, citing a lack of correlation between $\mu^{30}$Si and $\delta^{30}$Si across different meteorite groups. However, the effect is notably pronounced for enstatite chondrites. As discussed below, the difference of +11±3 in $\mu^{30}$Si between BSE and enstatite chondrites can be largely attributed to the fact that equilibrium between gas and solid fractionated the Si isotopic composition of these objects, which is not well accounted for by the exponential law used by Onyett *et al.* (2023) in their data reduction.

The $\delta^{30}$Si values of EH and EL are -0.68±0.03 and -0.55±0.04 ‰ respectively (Armytage et al., 2011; Dauphas et al., 2015; Fitoussi and Bourdon, 2012; Georg et al., 2007; Onyett et al., 2023; Savage and Moynier, 2013; Zambardi et al., 2013), while the silicate Earth is at -0.29±0.01 ‰ (Savage et al., 2010; Savage et al., 2014; Zambardi et al., 2013). These variations correlate with variations in Mg/Si ratios and are best explained by equilibrium isotopic fractionation between SiO gas and forsterite during condensation in the solar nebula (Dauphas et al., 2015). At that temperature, we expect a fractionation of ~+2.3 ‰ for $\delta^{30}$Si between forsterite and $SiO_g$, with approximately half of Si condensed and the remaining Si in the gas. High-temperature equilibrium isotopic fractionation should follow the relationship (Dauphas et al., 2022; Dauphas and Schauble, 2016a; Matsuhisa, Goldsmith and Clayton, 1978; Young, Galy and Nagahara, 2002),

$$\frac{\Delta\delta^{30}\mathrm{Si}_{\mathrm{eq}}}{\Delta\delta^{29}\mathrm{Si}_{\mathrm{eq}}} \simeq \frac{1/m_{28}-1/m_{30}}{1/m_{28}-1/m_{29}} \simeq 1.931, \qquad (6)$$

where $m_{28}$, $m_{29}$, and $m_{29}$ are the masses of the three Si isotopes. Onyett *et al.* (2023) adopted the exponential law to correct their data for mass fractionation by internal normalization, which assumes the following relationship (Dauphas and Schauble, 2016a),

$$\frac{\Delta\delta^{30}\mathrm{Si}_{\mathrm{exp}}}{\Delta\delta^{29}\mathrm{Si}_{\mathrm{eq}}} \simeq \frac{\ln(m_{30})-\ln(m_{28})}{\ln(m_{29})-\ln(m_{28})} \simeq 1.964. \qquad (7)$$

Correcting natural mass fractionation on the $^{30}$Si/$^{28}$Si ratio by internal normalization to a fixed $^{29}$Si/$^{28}$Si ratio assuming exponential mass fractionation (Eq. 7) can yield artifact anomalies when the actual fractionation follows a different law like the equilibrium law (Eq. 6). Tang and Dauphas (2012a) derived approximate analytic formulas to calculate apparent isotopic anomalies if an incorrect mass fractionation law is used for internal normalization. For $\mu^{30}$Si and $\delta^{30}$Si, it takes the form,

$$\mu^{30}\mathrm{Si} \simeq 500(n-k)\frac{(m_{30}-m_{29})}{m_{28}}\delta^{30}\mathrm{Si}, \qquad (8)$$

with $n$ and $k$ are parameters that describe the actual and inferred mass fractionation laws, respectively ($-1$ for high-T equilibrium and 0 for exponential; Maréchal, Télouk and Albarède, 1999)XXXX. If the exponential law ($k=0$) is used to correct high-T equilibrium isotopic fractionation (Tang and Dauphas, 2012a), then we expect the following relationship between apparent isotopic anomalies and mass fractionation, $\mu^{30}\mathrm{Si} \simeq -17\,\delta^{30}\mathrm{Si}$. Given the +0.4 ‰ difference in $\delta^{30}$Si, we would therefore expect silicate Earth to be shifted relative to enstatite chondrites by -7 ppm. Onyett *et al.* (2023) calculated a similar value for isotopic equilibrium involving gaseous atomic Si but reported different values for gas species coordinated by O or other atoms as they used isotopologue rather than isotope masses. Taking SiO as an example, they replaced $m_{28}$, $m_{29}$, and $m_{30}$ with $m_{28}+m_{\mathrm{O}}$, $m_{29}+m_{\mathrm{O}}$, and $m_{30}+m_{\mathrm{O}}$ ($m_{\mathrm{O}}$ is the mass of oxygen) in Eq. 6, corresponding to a slope $\Delta\delta^{30}\mathrm{Si}/\Delta\delta^{29}\mathrm{Si}$ of 1.954, which is closer to the exponential law (Extended Data Fig. 1 in Onyett *et al.* 2023). The high-temperature equilibrium mass fractionation law should however not depend on the atoms coordinating with Si (Dauphas and Schauble, 2016a; Matsuhisa, Goldsmith and Clayton, 1978; Young, Galy and Nagahara, 2002; see Appendix A1 of Dauphas *et al.* 2022 for a diatomic gas molecule). The equilibrium mass fractionation law calculated for $SiO_g$ by Onyett *et al*. (2023) is therefore incorrect. Correcting for this 7 ppm effect, the enstatite chondrite-silicate Earth difference shrinks from +11±3 to +4±3 (Fig. 10), which is barely resolvable. Other meteorite groups would also need to be corrected for the same effect. We conclude that much of the difference in $\mu^{30}$Si between the silicate Earth and enstatite chondrites is an artifact arising from correcting natural equilibrium isotopic fractionation in Si isotopes using the exponential law, which is inappropriate. This conclusion does not depend on details of the scenario outlined by Dauphas et al. (2015) being correct, only on high-temperature equilibrium controlling this fractionation, which is highly likely.

The two primary models for Earth's building blocks are as follows: (*i*) Earth was formed from material that, while isotopically akin to E-chondrites, possessed a distinct chemistry and a slightly different isotopic composition (Dauphas, 2017). (*ii*) Earth arose from a specific blend of achondrites and carbonaceous material. This blend underwent thermal processing, aligning certain isotopic anomalies (see Fig. 10) with those observed in both Earth and, coincidentally, enstatite chondrites (Schiller et al., 2020; Onyett et al., 2023). From a parsimony standpoint, the first model is favored. It demands fewer isotopic components, drawing primarily from enstatite rather than a blend of achondrites and carbonaceous materials. Additionally, it avoids invoking *ad hoc* nebular processes to alter isotopic anomalies at a large scale for refractory elements like Mo and Zr.

## 5. Conclusion

This study, along with numerous prior investigations over the past several years, has substantially enhanced the number and quality of isotopic anomaly measurements in planetary materials. We utilize these data from Mars and Earth in a Bayesian inference calculation to delineate the isotopic nature of their building blocks. The use of a Bayesian approach is motivated by its ability to resolve, in a statistically sound manner, the challenge of replicating Mars and Earth compositions via a mixing model while simultaneously permitting exploration of the isotopic compositions of the endmembers.

To simplify the problem, we apply principal component analysis and a weighted Euclidean distance metric to the database of isotopic anomaly measurements. Our results indicate that most variations documented in meteorites can be attributed to the mixing of CI, COCV, O, and E. Notably, the PCA analysis uncovers the isotopic trichotomy CI, CC, and NC, previously observed in cross-correlation diagrams.

Earth's isotopic compositions cannot be reproduced exactly by considering known meteorite groups only. With this caveat in mind, our analysis reveals that Earth consists of 92% $\mathcal{E}$, 6% $\mathcal{CI}$, and <2% $\mathcal{COCV}$ and $\mathcal{O}$. Mars' isotopic composition is best characterized as a blend of ~65% $\mathcal{E}$, 33% $\mathcal{O}$, and <2% $\mathcal{CI}$ and $\mathcal{COCV}$. There was a notable increase in the contribution of carbonaceous material in the last 40% of its accretion. Mars' accretion, on the other hand, saw a shift from an equal mixture of $\mathcal{E}$+$\mathcal{O}$ to predominantly $\mathcal{E}$. The nature of the late veneer remains poorly defined for both Mars and Earth, but it could have included a significant proportion of carbonaceous material.

We consider potential planetary dynamics scenarios that could have enabled the observed changes to take place. Mars may have originated at the boundary between the $\mathcal{O}$ and $\mathcal{E}$ regions, subsequently migrating into the $\mathcal{E}$ -region via gas-driven type-I migration over a span of a few million years. The increased contribution of carbonaceous material to Earth could be attributed to the random impact of an outer solar system embryo that migrated within the inner solar system while nebular gas remained, subsequently playing a role in the phase of chaotic growth.

Our study demonstrates that Si isotopic anomalies in Earth and enstatite chondrites should not be used as a constraint on the nature of the material accreted by Earth. These objects have highly fractionated Si isotopic compositions that, when inadequately corrected using the exponential law, produce apparent isotopic anomalies. These anomalies are not nucleosynthetic in origin; instead, they reflect nebular and planetary processes.

**Acknowledgements.** This work was supported by NASA grants 80NSSC20K1409 (Habitable Worlds), 359NNX17AE86G (LARS), NNX17AE87G and 80NSSC20K0821 (Emerging Worlds), EAR-2001098 (CSEDI), and a DOE grant to ND. Discussions with James Zhang were greatly appreciated. Editorial oversight by Dr. Morbidelli and comments by two anonymous reviewers were greatly appreciated.

**Author contributions.** ND and TH conceived the study. TH measured the Fe isotopic compositions. ND implemented the Bayesian inference calculation and ran the PCA/weighted Euclidean distance analysis. DN provided input on dynamical interpretations. ND wrote the first draft of the manuscript, which was subsequently edited by all authors.

**Appendix.** The Origins Lab isotopic anomaly database (OL_anomaly_database.xlsx) is publicly available as a *Google Sheets*. It will continue to be updated in the future by the authors as new measurements are published. It can be accessed at the following link:
https://drive.google.com/drive/folders/17IzyN4L7P8JyyG838-iq90wxgftltJ-k?usp=sharing

# Figures

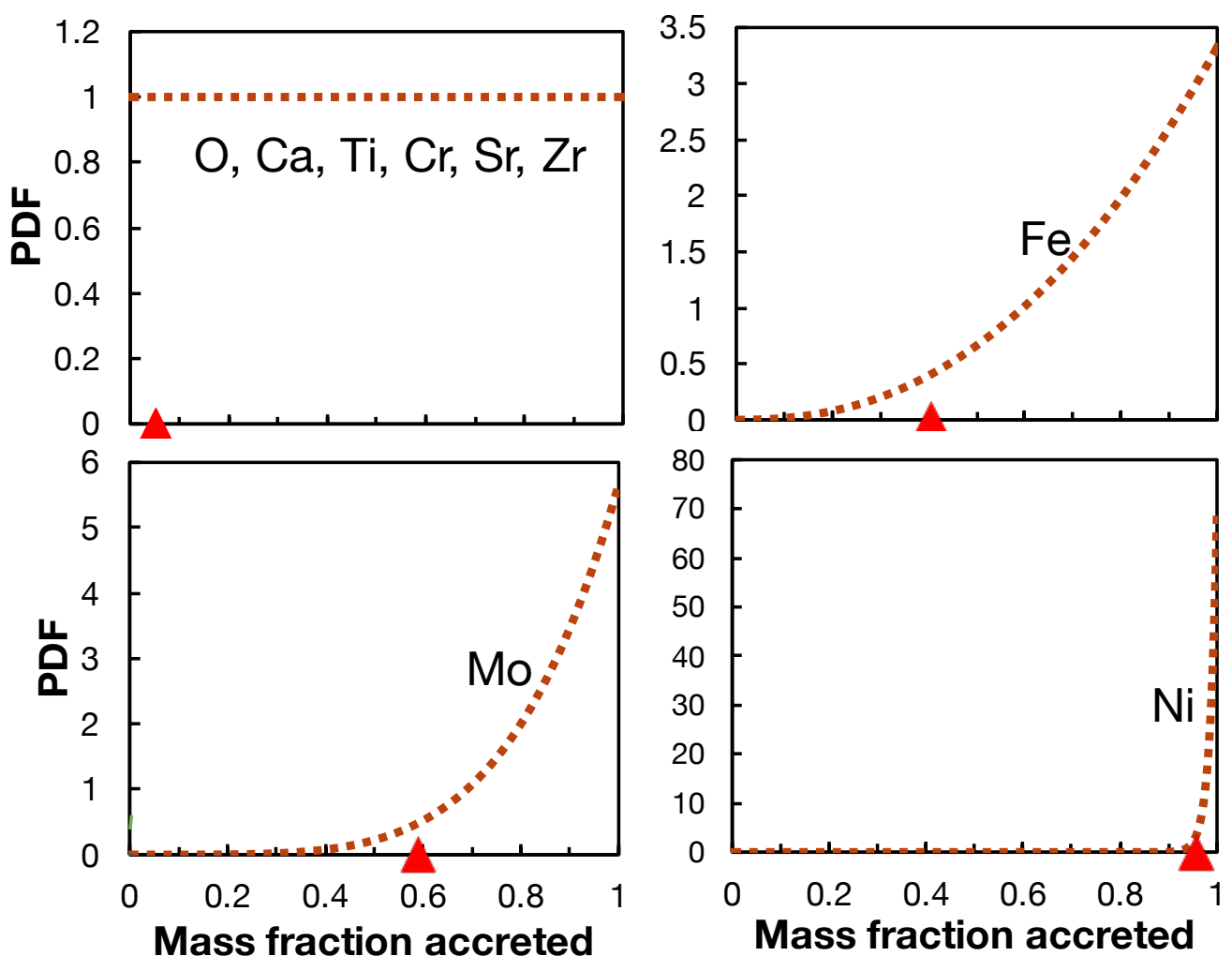

**Fig. 1.** Probability density function for the delivery of elements in Mars' mantle (Eq. 3; Dauphas, 2017). Due to pressure effects (Li and Agee, 1996), Ni is much less siderophile in Earth than in Mars. Nickel in the Martian mantle therefore records later accretion stages than in the terrestrial mantle.

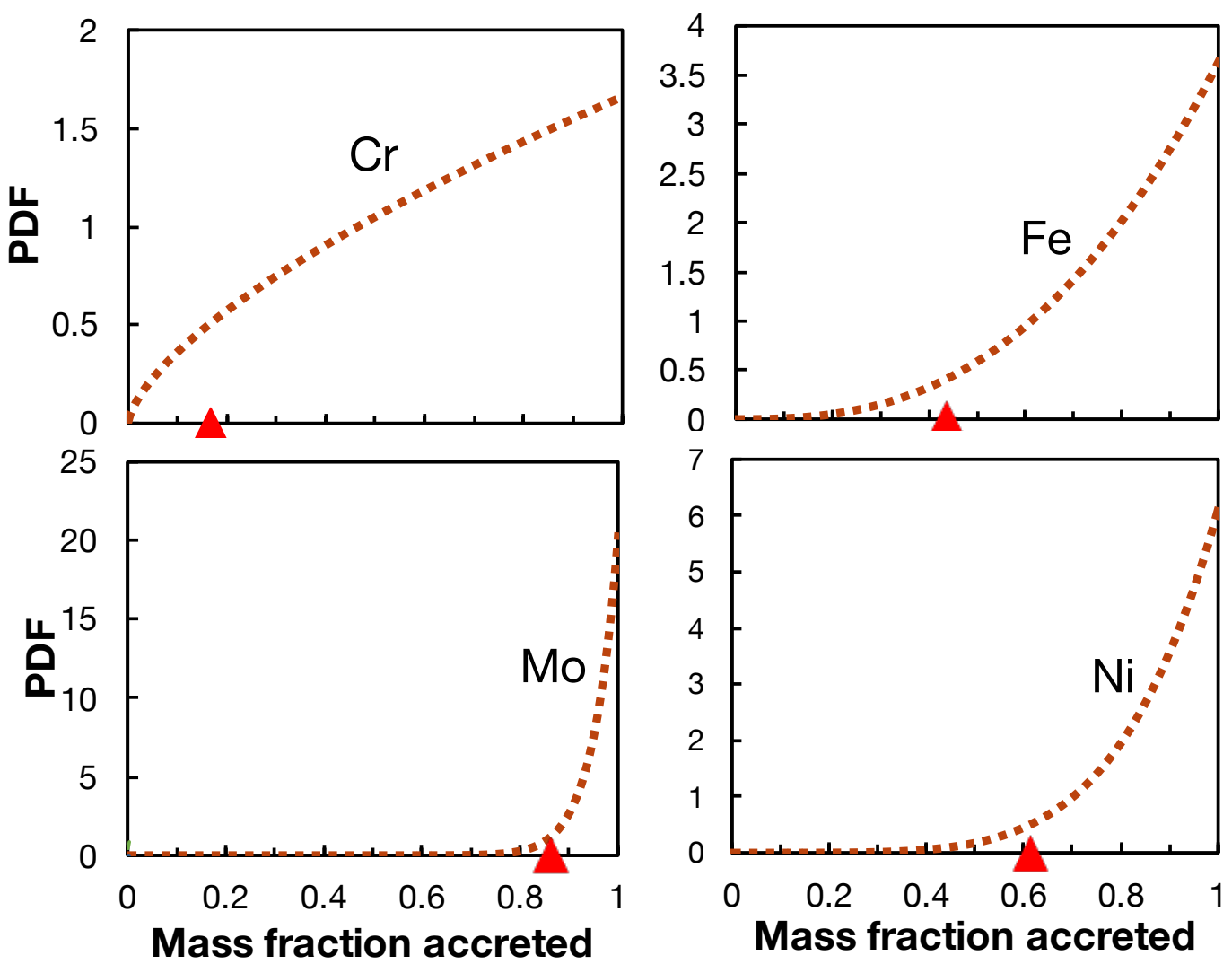

**Fig. 2.** Probability density function for the delivery of elements in Earth's mantle (Eq. 3; Dauphas, 2017).

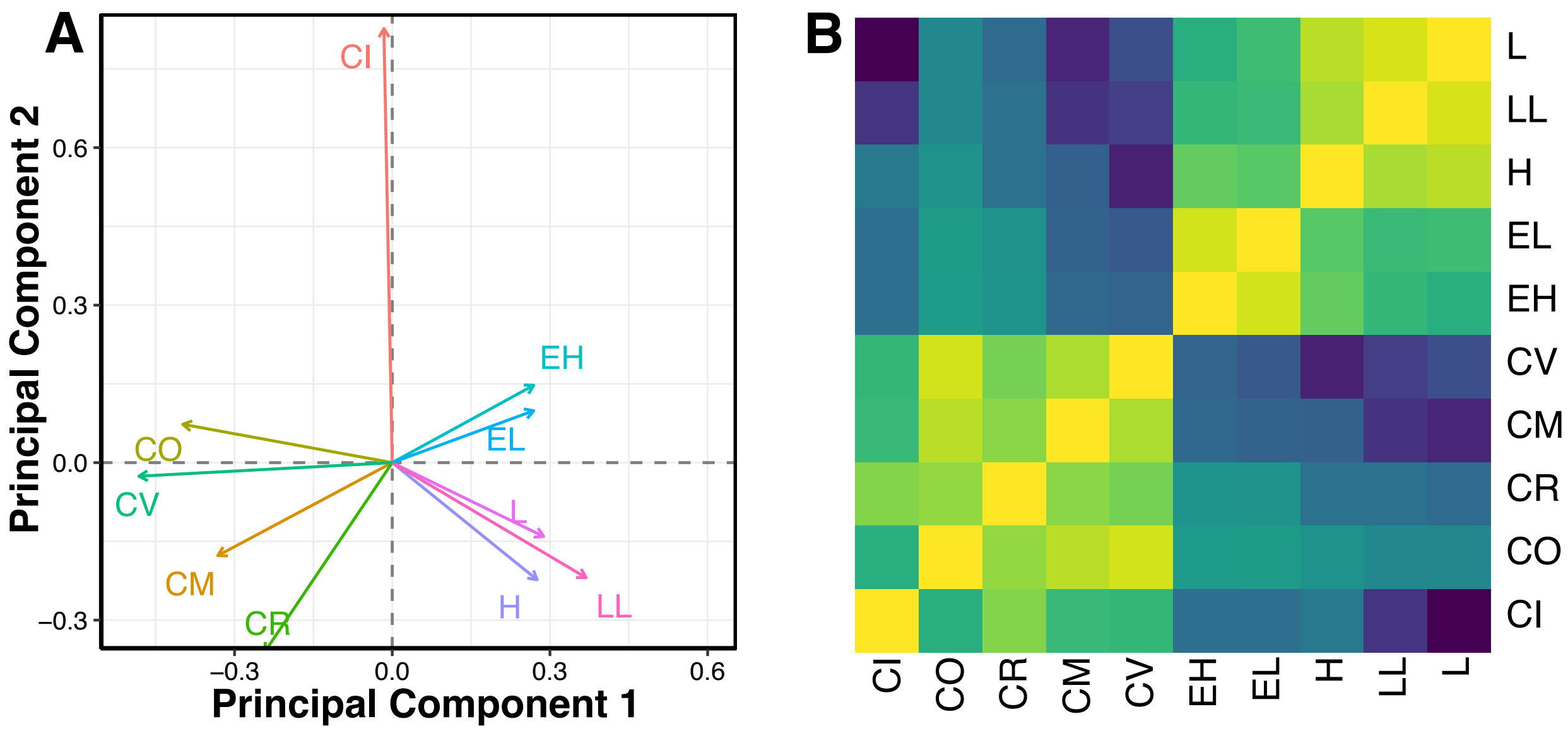

**Figure 3.** Taxonomy of chondrite groups (CI, CM, CO, CV, CR, EH, EL, H, L, LL) based on isotopic anomalies ($\Delta^{17}$O, $\varepsilon^{48}$Ca, $\varepsilon^{50}$Ti, $\varepsilon^{54}$Cr, $\varepsilon^{54}$Fe, $\varepsilon^{64}$Ni, $\mu^{66}$Zn, $\mu^{84}$Sr $\varepsilon^{96}$Zr, $\varepsilon^{92}$Mo, $\Delta^{95}$Mo, $\varepsilon^{100}$Ru). These meteorite groups are the only ones among undifferentiated meteorites with a complete dataset of isotopic anomalies (Table 4). **A.** PCA loadings diagram for meteorite groups, which are treated as variables, with isotopic anomalies as the samples. The aim is to reduce the dimensionality of meteorite groups rather than isotopic anomalies. Since the isotopic anomalies are measured on disparate scales and span a variety of ranges, we first normalized them using z-score normalization. The direction and magnitude of each arrow in the loadings plot indicates the contribution of each meteorite group to the respective principal component. The PCA analysis allows for a clear separation of meteorite groups into the trichotomy NC=EH+EL+H+L+LL, CC=CO+CV+CM+CR, and CI. The analysis further illuminates the divide between ordinary (O=H+L+LL) and enstatite (E=EH+EL) chondrites. **B.** Weighted Euclidean distance matrix depicting the distances between meteorite groups in isotopic anomaly space. The distances were calculated by weighting each difference by the inverse of its variance, thereby ensuring that less precise measurements contribute less to the distance metrics. For two isotopic vectors $i1$ and $i2$, each associated with standard deviation vectors $\sigma_{i1}$ and $\sigma_{i2}$, the weighted Euclidean distance is computed as $d_{i2-i1} = \sqrt{\sum_{j=1}^{n_j} \left(I_{i2,j} - I_{i2,j}\right)^2 / \left(\sigma_{i2,j}^2 + \sigma_{i1,j}^2\right)}$. This calculation ensures that the effects of less certain measurements are minimized. Like the PCA, the distance matrix illuminates several readily identifiable subgroups within the data.

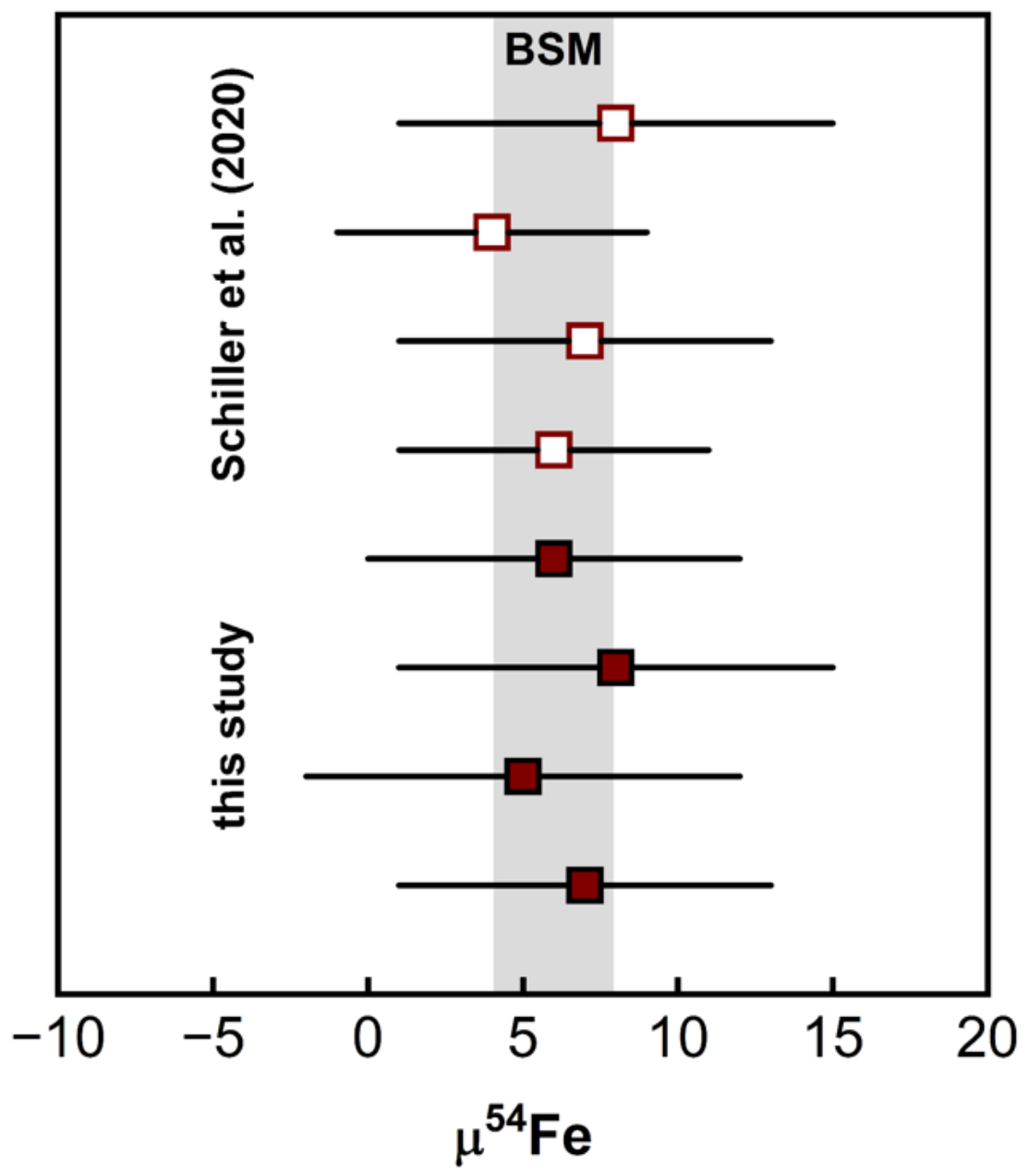

**Fig. 4.** Fe isotopic composition of Martian meteorites. The four samples that represent the three major groups of Martian meteorites (shergottites, nakhlites, and chassignites) have indistinguishable $\mu^{54}$Fe values that agree with available data of Schiller *et al*. (2020). Together these samples define a weighted average for the bulk silicate Mars (BSM) of +7±2 (95% c.i.).

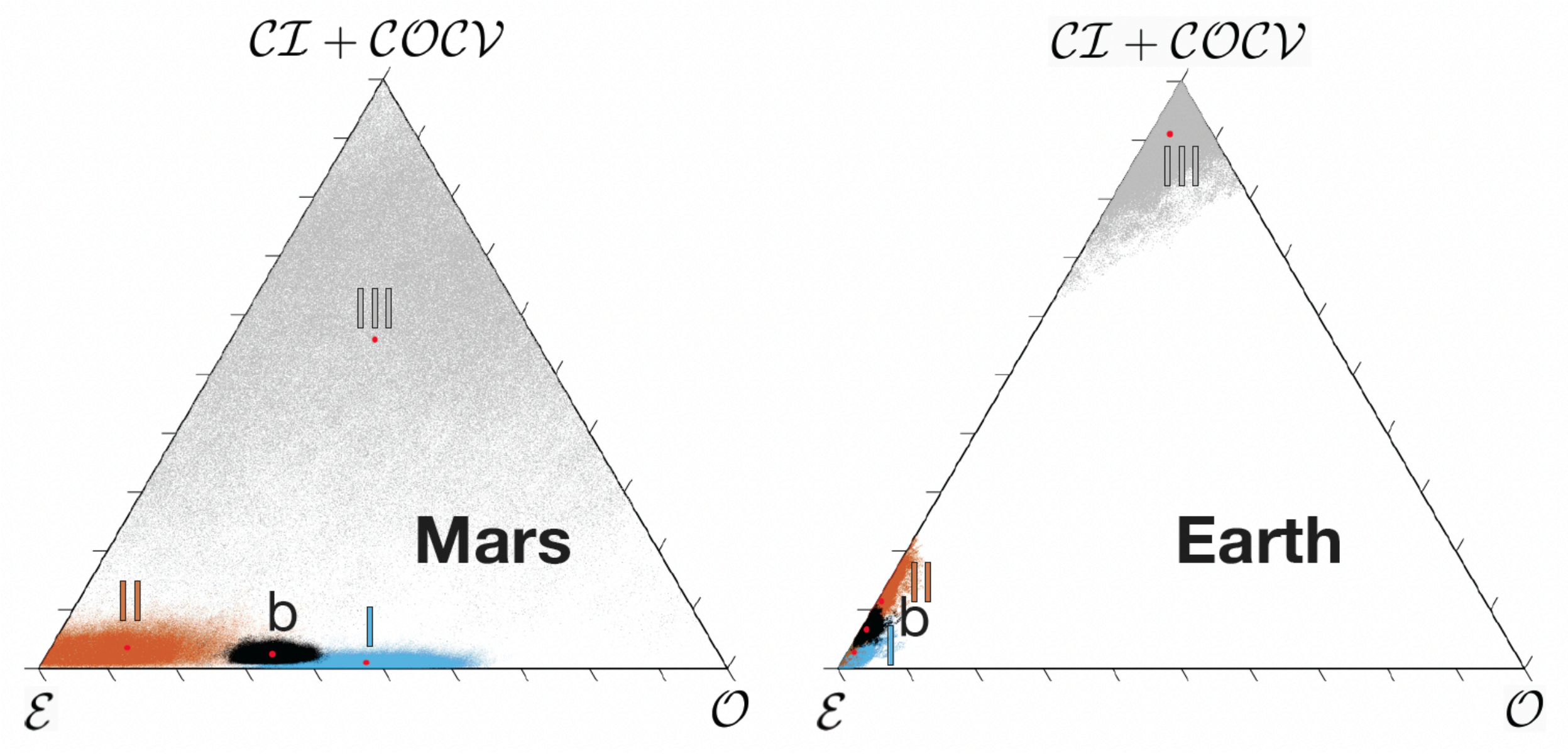

**Fig. 5.** Contributions of $\mathcal{O}$, $\mathcal{E}$, and carbonaceous asteroids to accretion stages of Mars and Earth in the baseline model (Tables 7, 8; III is the late veneer; II=60-99.x%, I=0-60%).

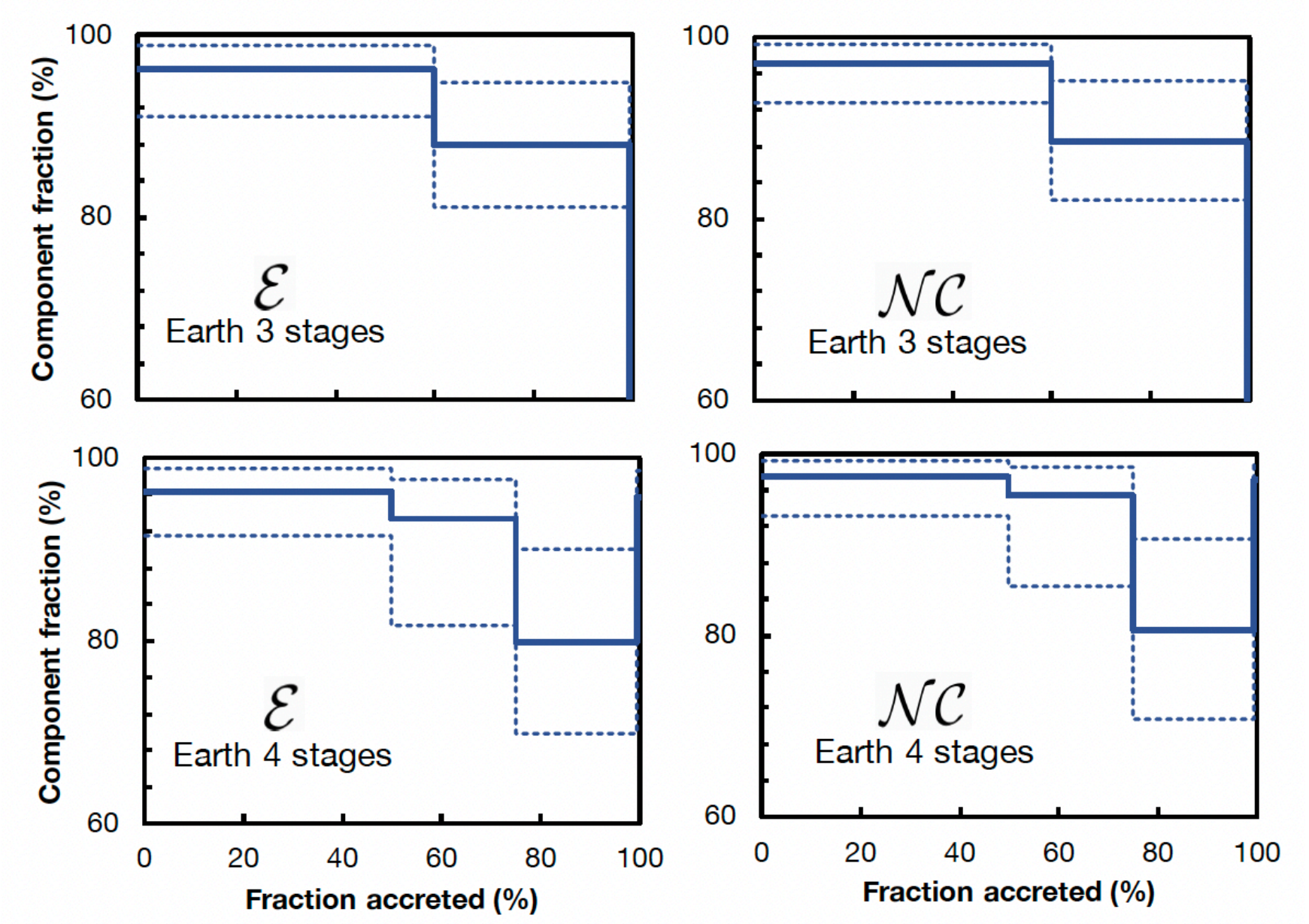

**Fig. 6.** Contributions of $\mathcal{E}$ and $\mathcal{NC}$ during accretion of Earth in the *baseline* and *4-stage* models.

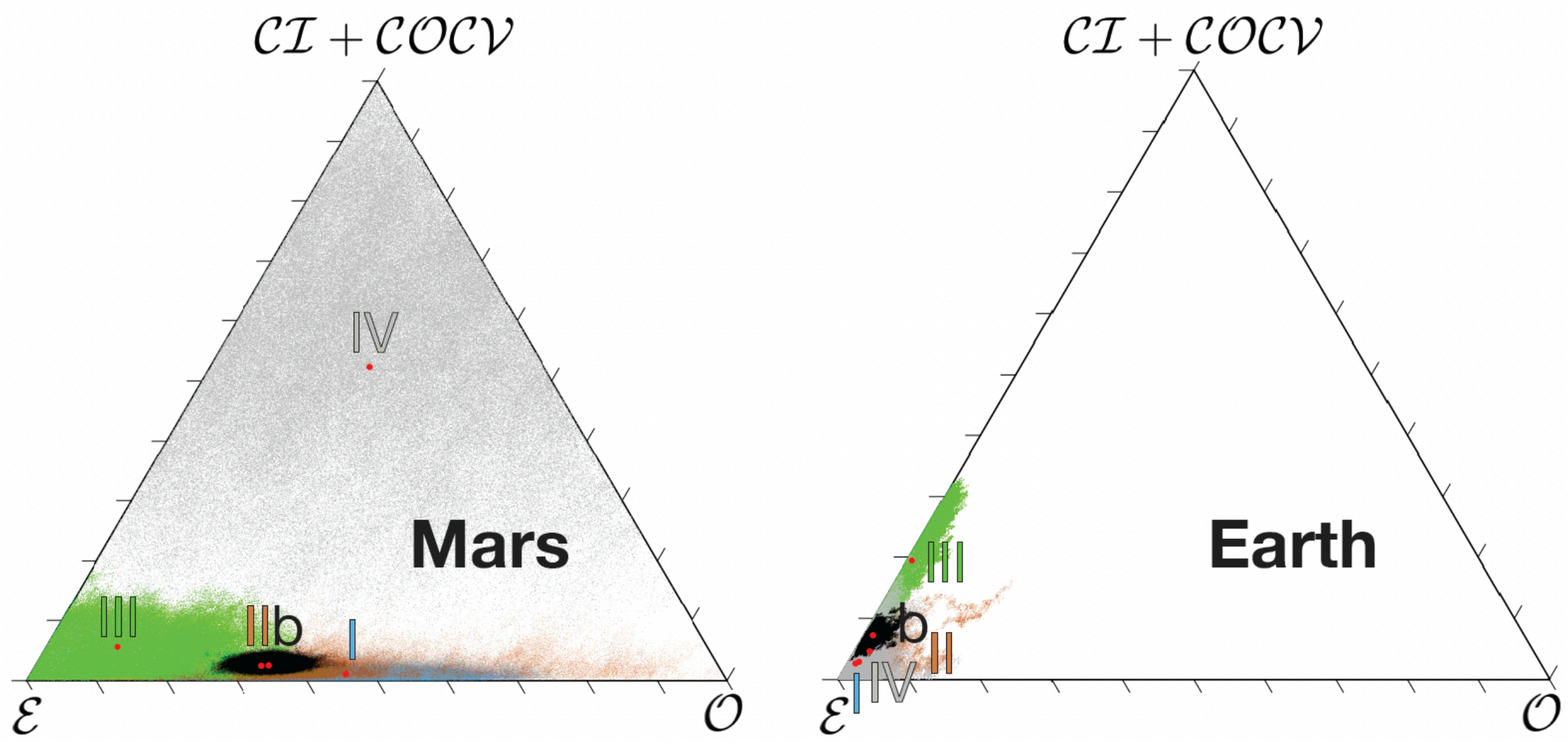

**Fig. 7.** Contributions of $\mathcal{O}$, $\mathcal{E}$, and carbonaceous asteroids to accretion stages of Mars and Earth in the *4-stage* model (Tables 7, 8; IV is the late veneer; III=75-99.x%, II=50-75%, I=0-50%).

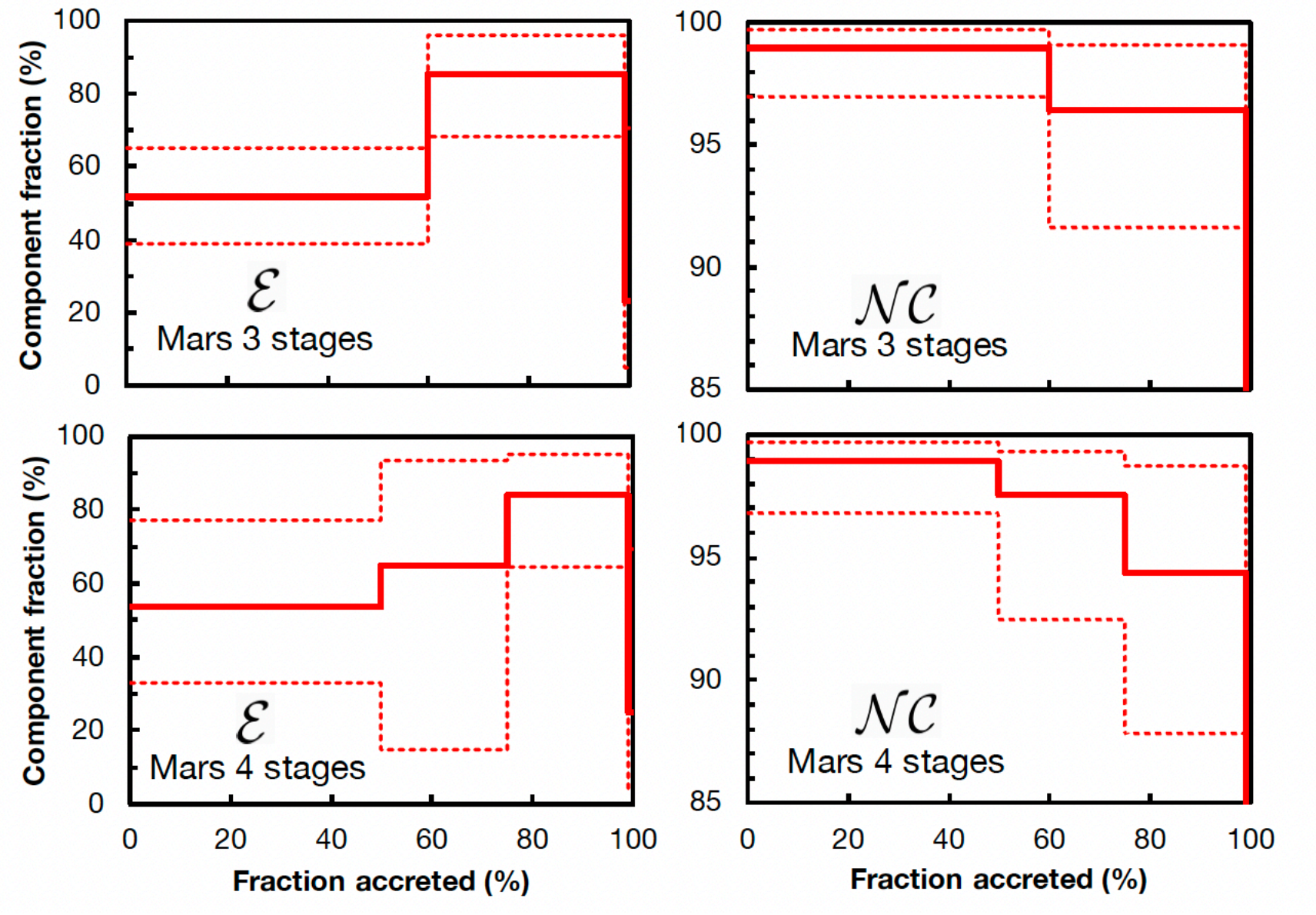

**Fig. 8.** Contributions of $\mathcal{E}$ and $\mathcal{NC}$ during accretion of Mars in the *baseline* and *4-stage* models.

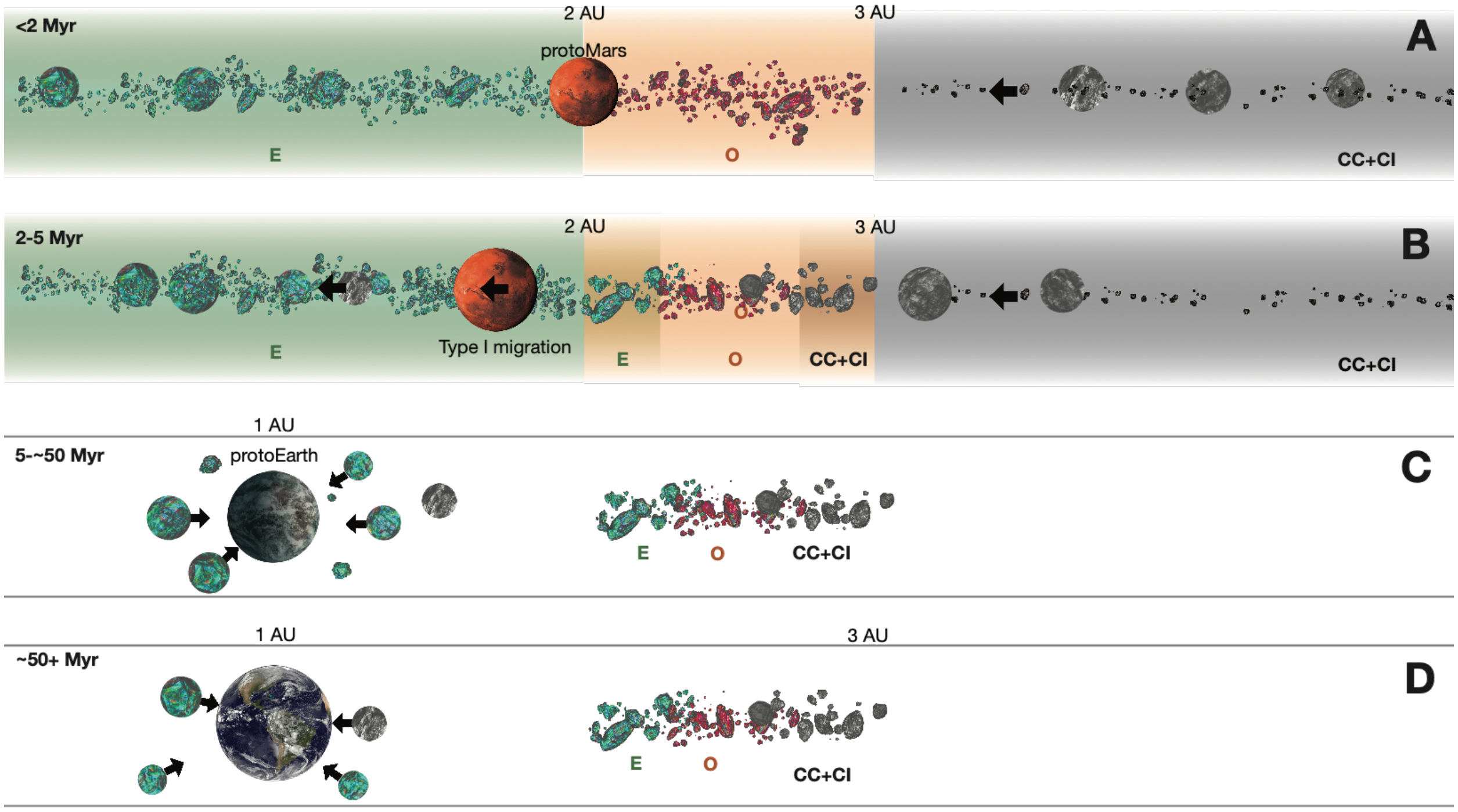

**Fig. 9.** Schematic illustration of transport and mixing between different isotopic reservoirs during growth of Mars and Earth. Mars was born near the inner edge of the main asteroid belt, in between the $\mathcal{E}$ and $\mathcal{O}$ reservoirs. It started accreted an equal mixture of both planetesimals (**A**) but progressively accreted more $\mathcal{E}$ material as it migrated inwards into the $\mathcal{E}$ region by type I migration (**B**). During those times, carbonaceous asteroids from the outer solar system were scattered inwards by the giant planets, but interaction with nebular gas led to their implantation into the outer main belt. Larger carbonaceous objects were less subjected to gas drag and were able to penetrate in the inner solar system. Earth grew later primarily from $\mathcal{E}$ -type embryos (**C**). Several tens of millions of years after solar system formation, one or several interloper carbonaceous embryos implanted in the inner disk during the first phase were stochastically accreted by the Earth (**D**).

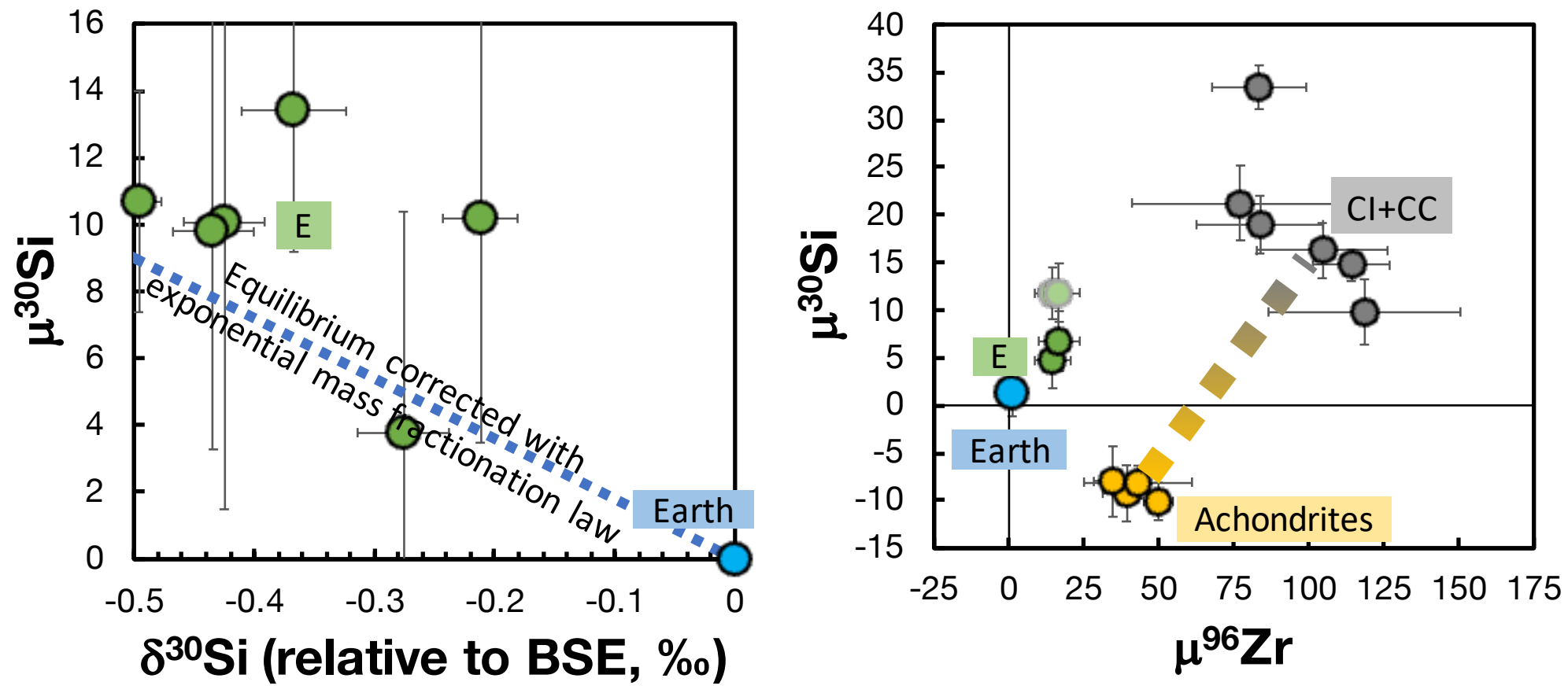

**Fig. 10.** Silicon isotopic relationships between silicate Earth and enstatite chondrites. **Left panel**: enstatite chondrites have fractionated $\delta^{30}$Si values relative to Earth (Fitoussi and Bourdon, 2012; Onyett et al., 2023). This is most likely due to high-temperature equilibrium Si isotopic fractionation in the nebula (Dauphas et al., 2015). Correction of such equilibrium isotopic fractionation using the exponential law will yield spurious isotopic anomaly like those measured in enstatite chondrites (Eq. 6; Tang and Dauphas, 2012a). Much of the positive $\mu^{30}$Si values in enstatite chondrites relative to Earth (Onyett et al., 2023) can be explained by such an artifact of the internal normalization procedure (see text for details; data points in this panel correspond to individual $\delta^{30}$Si-$\mu^{30}$Si meteorite analyses from Onyett *et al.* 2023). The semi-transparent green circles are the raw values for enstatite chondrites, and the solid green circles are corrected for equilibrium isotopic fractionation (left panel). **Right panel**: Si and Zr isotopic anomalies in the silicate Earth, enstatite chondrites, non-aubrite achondrites, and carbonaceous chondrites (Table 4). Mixing achondrites and carbonaceous chondrites can emulate the $\mu^{30}$Si of Earth but yields large $\mu^{96}$Zr anomalies that are not seen in Earth. After correction for equilibrium Si isotopic fractionation, enstatite chondrites have $\mu^{30}$Si close to Earth, with no need to invoke multicomponent mixing and hypothetical fractionation of presolar SiC (Onyett et al., 2023).

## References

Adachi, I., Hayashi, C., Nakazawa, K., 1976. The gas drag effect on the elliptic motion of a solid body in the primordial solar nebula. Progress of Theoretical Physics. 56, 1756-1771.

Allègre, C., Manhès, G., Lewin, É., 2001. Chemical composition of the Earth and the volatility control on planetary genetics. Earth and Planetary Science Letters. 185, 49-69.

Allegre, C. J., Poirier, J.-P., Humler, E., Hofmann, A. W., 1995. The chemical composition of the Earth. Earth and Planetary Science Letters. 134, 515-526.

Armytage, R., Georg, R., Savage, P., Williams, H., Halliday, A., 2011. Silicon isotopes in meteorites and planetary core formation. Geochimica et Cosmochimica Acta. 75, 3662-3676.

Brasser, R., Dauphas, N., Mojzsis, S., 2018. Jupiter's influence on the building blocks of Mars and earth. Geophysical Research Letters. 45, 5908-5917.

Brož, M., Chrenko, O., Nesvorný, D., Dauphas, N., 2021. Early terrestrial planet formation by torque-driven convergent migration of planetary embryos. Nature Astronomy. 5, 898-902.

Budde, G., Burkhardt, C., Brennecka, G. A., Fischer-Gödde, M., Kruijer, T. S., Kleine, T., 2016. Molybdenum isotopic evidence for the origin of chondrules and a distinct genetic heritage of carbonaceous and non-carbonaceous meteorites. Earth and Planetary Science Letters. 454, 293-303.

Budde, G., Burkhardt, C., Kleine, T., 2019. Molybdenum isotopic evidence for the late accretion of outer Solar System material to Earth. Nature Astronomy. 3, 736-741.

Budde, G., Tissot, F. L., Kleine, T., Marquez, R. T., 2023. Spurious molybdenum isotope anomalies resulting from non-exponential mass fractionation. Geochemistry. 126007.

Burkhardt, C., Kleine, T., Oberli, F., Pack, A., Bourdon, B., Wieler, R., 2011a. Molybdenum isotope anomalies in meteorites: Constraints on solar nebula evolution and origin of the Earth. Earth and Planetary Science Letters. 312, 390-400.

Burkhardt, C., Kleine, T., Oberli, F., Pack, A., Bourdon, B., Wieler, R., 2011b. Molybdenum isotope anomalies in meteorites: Constraints on solar nebula evolution and origin of the Earth. Earth and Planetary Science Letters. 312, 390-400.

Burkhardt, C., Spitzer, F., Morbidelli, A., Budde, G., Render, J. H., Kruijer, T. S., Kleine, T., 2021. Terrestrial planet formation from lost inner solar system material. Science advances. 7, eabj7601.

Chambers, J. E., Wetherill, G. W., 1998. Making the Terrestrial Planets: N-Body Integrations of Planetary Embryos in Three Dimensions. Icarus. 136, 304-327.

Chib, S., Greenberg, E., 1995. Understanding the metropolis-hastings algorithm. The american statistician. 49, 327-335.

Clayton, R. N., 2003. Oxygen isotopes in the solar system. Solar System History from Isotopic Signatures of Volatile Elements: Volume Resulting from an ISSI Workshop 14–18 January 2002, Bern, Switzerland. Springer, pp. 19-32.

Clayton, R. N., Mayeda, T. K., 1996. Oxygen isotope studies of achondrites. Geochimica et Cosmochimica Acta. 60, 1999-2017.

Clement, M. S., Kaib, N. A., Raymond, S. N., Walsh, K. J., 2018. Mars' growth stunted by an early giant planet instability. Icarus. 311, 340-356.

Dauphas, N., 2017. The isotopic nature of the Earth's accreting material through time. Nature. 541, 521-524.

Dauphas, N., Chaussidon, M., 2011. A perspective from extinct radionuclides on a young stellar object: The Sun and its accretion disk. Annual Review of Earth and Planetary Sciences. 39, 351-386.

Dauphas, N., Chen, J. H., Zhang, J., Papanastassiou, D. A., Davis, A. M., Travaglio, C., 2014a. Calcium-48 isotopic anomalies in bulk chondrites and achondrites: evidence for a uniform isotopic reservoir in the inner protoplanetary disk. Earth and Planetary Science Letters. 407, 96-108.

Dauphas, N., Chen, J. H., Zhang, J., Papanastassiou, D. A., Davis, A. M., Travaglio, C., 2014b. Calcium-48 isotopic anomalies in bulk chondrites and achondrites: evidence for a uniform isotopic reservoir in the inner protoplanetary disk. Submitted to Earth and Planetary Science Letters.

Dauphas, N., Marty, B., Reisberg, L., 2002a. Molybdenum evidence for inherited planetary scale isotope heterogeneity of the protosolar nebula. Astrophysical Journal. 565, 640-644.

Dauphas, N., Marty, B., Reisberg, L., 2002b. Molybdenum nucleosynthetic dichotomy revealed in primitive meteorites. Astrophysical Journal. 569, L139-L142.

Dauphas, N., et al., 2022. The extent, nature, and origin of K and Rb depletions and isotopic fractionations in Earth, the Moon, and other planetary bodies. The Planetary Science Journal. 3, 29.

Dauphas, N., Poitrasson, F., Burkhardt, C., Kobayashi, H., Kurosawa, K., 2015. Planetary and meteoritic Mg/Si and δ30Si variations inherited from solar nebula chemistry. Earth and Planetary Science Letters. 427, 236-248.

Dauphas, N., Pourmand, A., 2011a. Hf-W-Th evidence for rapid growth of Mars and its status as a planetary embryo. Nature. 473, 489-492.

Dauphas, N., Pourmand, A., 2011b. Hf–W–Th evidence for rapid growth of Mars and its status as a planetary embryo. Nature. 473, 489-492.

Dauphas, N., Schauble, E., 2016a. Mass fractionation laws, mass-independent effects, and isotopic anomalies. Annual Review of Earth and Planetary Sciences. in press.

Dauphas, N., Schauble, E. A., 2016b. Mass fractionation laws, mass-independent effects, and isotopic anomalies. Annual Review of Earth and Planetary Sciences. 44, 709-783.

Davis, A. M., Richter, F. M., Mendybaev, R. A., Janney, P. E., Wadhwa, M., McKeegan, K. D., 2015. Isotopic mass fractionation laws for magnesium and their effects on 26Al–26Mg systematics in solar system materials. Geochimica et Cosmochimica Acta. 158, 245-261.

Day, J. M., Brandon, A. D., Walker, R. J., 2016. Highly siderophile elements in Earth, Mars, the Moon, and asteroids. Reviews in Mineralogy and Geochemistry. 81, 161-238.

Desch, S. J., Kalyaan, A., Alexander, C. M. D., 2018. The effect of Jupiter's formation on the distribution of refractory elements and inclusions in meteorites. The Astrophysical Journal Supplement Series. 238, 11.

Fischer-Gödde, M., Burkhardt, C., Kruijer, T. S., Kleine, T., 2015. Ru isotope heterogeneity in the solar protoplanetary disk. Geochimica et Cosmochimica Acta. 168, 151-171.

Fischer-Gödde, M., et al., 2020. Ruthenium isotope vestige of Earth's pre-late-veneer mantle preserved in Archaean rocks. Nature. 579, 240-244.

Fischer-Gödde, M., Kleine, T., 2017. Ruthenium isotopic evidence for an inner Solar System origin of the late veneer. Nature. 541, 525-527.

Fitoussi, C., Bourdon, B., 2012. Silicon isotope evidence against an enstatite chondrite Earth. Science. 335, 1477-1480.

Fitoussi, C., Bourdon, B., Wang, X., 2016. The building blocks of Earth and Mars: A close genetic link. Earth and Planetary Science Letters. 434, 151-160.

Gelfand, A. E., Smith, A. F., 1990. Sampling-based approaches to calculating marginal densities. Journal of the American statistical association. 85, 398-409.

Georg, R. B., Halliday, A. N., Schauble, E. A., Reynolds, B. C., 2007. Silicon in the Earth’s core. Nature. 447, 1102-1106.

Gladman, B. J., et al., 1997. Dynamical lifetimes of objects injected into asteroid belt resonances. Science. 277, 197-201.

Goldreich, P., Tremaine, S., 1980. Disk-satellite interactions. Astrophysical Journal. 241, 425-441.

Gomes, R., Levison, H. F., Tsiganis, K., Morbidelli, A., 2005. Origin of the cataclysmic Late Heavy Bombardment period of the terrestrial planets. Nature. 435, 466-469.

Gu, J. T., et al., 2023. Comparisons of the core and mantle compositions of earth analogs from different terrestrial planet formation scenarios. Icarus. 115425.

Hart, S. R., Zindler, A., 1986. In search of a bulk-Earth composition. Chemical Geology. 57, 247-267.

Hastings, W. K., 1970. Monte Carlo sampling methods using Markov chains and their applications.

Hopp, T., et al., 2022a. Ryugu’s nucleosynthetic heritage from the outskirts of the Solar System. Science Advances. 8, eadd8141.

Hopp, T., Dauphas, N., Spitzer, F., Burkhardt, C., Kleine, T., 2022b. Earth's accretion inferred from iron isotopic anomalies of supernova nuclear statistical equilibrium origin. Earth and Planetary Science Letters. 577, 117245.

Jagoutz, E., et al., 1979. The abundances of major, minor and trace elements in the earth's mantle as derived from primitive ultramafic nodules. In: Lunar and Planetary Science Conference, 10th, Houston, Tex., March 19-23, 1979, Proceedings. Volume 2.(A80-23617 08-91) New York, Pergamon Press, Inc., 1979, p. 2031-2050. Research supported by the Deutsche Forschungsgemeinschaft., Vol. 10, pp. 2031-2050.

Javoy, M., 1995. The integral enstatite chondrite model of the Earth. Geophysical Research Letters. 22, 2219-2222.

Javoy, M., 1997. The major volatile elements of the Earth: Their origin, behavior, and fate. Geophysical Research Letters. 24, 177-180.

Javoy, M., et al., 2010a. The chemical composition of the Earth: Enstatite chondrite models. Earth and Planetary Science Letters. 293, 259-268.

Javoy, M., et al., 2010b. The chemical composition of the Earth: enstatite chondrite models. Earth and Planetary Science Letters. 293, 259-268.

Johansen, A., Ronnet, T., Bizzarro, M., Schiller, M., Lambrechts, M., Nordlund, Å., Lammer, H., 2021. A pebble accretion model for the formation of the terrestrial planets in the Solar System. Science Advances. 7, eabc0444.

Kipp, M. A., Tissot, F. L., 2022. Inverse methods for consistent quantification of seafloor anoxia using uranium isotope data from marine sediments. Earth and Planetary Science Letters. 577, 117240.

Kleine, T., Budde, G., Burkhardt, C., Kruijer, T., Worsham, E., Morbidelli, A., Nimmo, F., 2020. The non-carbonaceous–carbonaceous meteorite dichotomy. Space Science Reviews. 216, 1-27.

Kleine, T., Steller, T., Burkhardt, C., Nimmo, F., 2023. An inner solar system origin of volatile elements in Mars. Icarus. 397, 115519.

Kobayashi, H., Dauphas, N., 2013. Small planetesimals in a massive disk formed Mars. Icarus.

Komaki, A., Fukuhara, S., Suzuki, T. K., Yoshida, N., 2023. Simulations of Protoplanetary Disk Dispersal: Stellar Mass Dependence of the Disk Lifetime. arXiv preprint arXiv:2304.13316.

Krot, A. N., Amelin, Y., Cassen, P., Meibom, A., 2005. Young chondrules in CB chondrites from a giant impact in the early solar system. Nature. 436, 989-992.

Kruijer, T. S., Burkhardt, C., Budde, G., Kleine, T., 2017. Age of Jupiter inferred from the distinct genetics and formation times of meteorites. Proceedings of the National Academy of Sciences. 114, 6712-6716.

Levison, H. F., Kretke, K. A., Walsh, K. J., Bottke, W. F., 2015. Growing the terrestrial planets from the gradual accumulation of submeter-sized objects. Proceedings of the National Academy of Sciences. 112, 14180-14185.

Li, J., Agee, C. B., 1996. Geochemistry of mantle–core differentiation at high pressure. Nature. 381, 686-689.

Liu, B., Liang, Y., 2017. An introduction of Markov chain Monte Carlo method to geochemical inverse problems: Reading melting parameters from REE abundances in abyssal peridotites. Geochimica et Cosmochimica Acta. 203, 216-234.

Lodders, K., 2021. Relative Atomic Solar System Abundances, Mass Fractions, and Atomic Masses of the Elements and Their Isotopes, Composition of the Solar Photosphere, and Compositions of the Major Chondritic Meteorite Groups. Space Science Reviews. 217, 44.

Lodders, K., Fegley, B., 1997. An Oxygen Isotope Model for the Composition of Mars. Icarus. 126, 373-394.

Maréchal, C. N., Télouk, P., Albarède, F., 1999. Precise analysis of copper and zinc isotopic compositions by plasma-source mass spectrometry. Chemical Geology. 156, 251-273.

Martins, R., Kuthning, S., Coles, B. J., Kreissig, K., Rehkämper, M., 2023. Nucleosynthetic isotope anomalies of zinc in meteorites constrain the origin of Earth's volatiles. Science. 379, 369-372.

Matsuhisa, Y., Goldsmith, J. R., Clayton, R. N., 1978. Mechanisms of hydrothermal crystallization of quartz at 250 C and 15 kbar. Geochimica et Cosmochimica Acta. 42, 173-182.

McCord, T. B., Adams, J. B., Johnson, T. V., 1970. Asteroid Vesta: Spectral reflectivity and compositional implications. Science. 168, 1445-1447.

McDonough, W., 2003. 3.16–Compositional model for the Earth's core. Treatise on geochemistry. 2, 547-568.

McDonough, W. F., Sun, S.-S., 1995. The composition of the Earth. Chemical Geology. 120, 223-253.

McNeil, D., Duncan, M., Levison, H. F., 2005. Effects of type I migration on terrestrial planet formation. The Astronomical Journal. 130, 2884.

McSween Jr, H. Y., et al., 2013. Dawn; the Vesta–HED connection; and the geologic context for eucrites, diogenites, and howardites. Meteoritics & Planetary Science. 48, 2090-2104.

Metropolis, N., Rosenbluth, A. W., Rosenbluth, M. N., Teller, A. H., Teller, E., 1953. Equation of state calculations by fast computing machines. The journal of chemical physics. 21, 1087-1092.

Morbidelli, A., Chambers, J., Lunine, J., Petit, J.-M., Robert, F., Valsecchi, G. B., Cyr, K., 2000. Source regions and timescales for the delivery of water to the Earth. Meteoritics & Planetary Science. 35, 1309-1320.

Morbidelli, A., Libourel, G., Palme, H., Jacobson, S. A., Rubie, D. C., 2020. Subsolar Al/Si and Mg/Si ratios of non-carbonaceous chondrites reveal planetesimal formation during early condensation in the protoplanetary disk. Earth and Planetary Science Letters. 538, 116220.

Morbidelli, A., Lunine, J. I., O'Brien, D. P., Raymond, S. N., Walsh, K. J., 2012. Building terrestrial planets. Annual Review of Earth and Planetary Sciences. 40, 251-275.

Nesvorny, D., Dauphas, N., Vokrouhlicky, D., Deienno, R., Hopp, T., 2023. Isotopic trichotomy of main belt asteroids from implantation of outer solar system planetesimals. Earth and Planetary Science Letters. submitted.

Nesvorný, D., Morbidelli, A., 2012. Statistical study of the early solar system's instability with four, five, and six giant planets. The Astronomical Journal. 144, 117.

Nesvorný, D., Roig, F. V., Deienno, R., 2021. The role of early giant-planet instability in terrestrial planet formation. The Astronomical Journal. 161, 50.

Nie, N. X., Wang, D., Torrano, Z. A., Carlson, R. W., O'D. Alexander, C. M., Shahar, A., 2023. Meteorites have inherited nucleosynthetic anomalies of potassium-40 produced in supernovae. Science. 379, 372-376.

Nimmo, F., O'brien, D., Kleine, T., 2010. Tungsten isotopic evolution during late-stage accretion: constraints on Earth–Moon equilibration. Earth and Planetary Science Letters. 292, 363-370.

O'Neill, H., 2014 3.1 Cosmochemical estimates of mantle composition. Treatise on Geochemistry: Oxford. Elsevier, pp. 1-39.

Onyett, I. J., Schiller, M., Makhatadze, G. V., Deng, Z., Johansen, A., Bizzarro, M., 2023. Silicon isotope constraints on terrestrial planet accretion. Nature. 1-6.

Paquet, M., et al., 2023. Contribution of Ryugu-like material to Earth's volatile inventory by Cu and Zn isotopic analysis. Nature astronomy. 7, 182-189.

Paquet, M., Sossi, P. A., Moynier, F., 2023. Origin and abundances of volatiles on Mars from the zinc isotopic composition of Martian meteorites. Earth and Planetary Science Letters. 611, 118126.

Pringle, E. A., Moynier, F., Savage, P. S., Badro, J., Barrat, J.-A., 2014. Silicon isotopes in angrites and volatile loss in planetesimals. Proceedings of the National Academy of Sciences. 111, 17029-17032.

Ptáček, M. P., Dauphas, N., Greber, N. D., 2020. Chemical evolution of the continental crust from a data-driven inversion of terrigenous sediment compositions. Earth and planetary science letters. 539, 116090.

Qin, L., Carlson, R. W., 2016. Nucleosynthetic isotope anomalies and their cosmochemical significance. Geochemical Journal. 50, 43-65.

Raymond, S. N., Izidoro, A., 2017. Origin of water in the inner Solar System: Planetesimals scattered inward during Jupiter and Saturn's rapid gas accretion. Icarus. 297, 134-148.

Render, J., Brennecka, G. A., 2021. Isotopic signatures as tools to reconstruct the primordial architecture of the Solar System. Earth and Planetary Science Letters. 555, 116705.
Render, J., Brennecka, G. A., Burkhardt, C., Kleine, T., 2022. Solar System evolution and terrestrial planet accretion determined by Zr isotopic signatures of meteorites. Earth and Planetary Science Letters. 595, 117748.
Rubie, D. C., et al., 2015a. Accretion and differentiation of the terrestrial planets with implications for the compositions of early-formed Solar System bodies and accretion of water. Icarus. 248, 89-108.
Rubie, D. C., et al., 2015b. Accretion and differentiation of the terrestrial planets with implications for the compositions of early-formed Solar System bodies and accretion of water. Icarus. 248, 89-108.
Rudge, J. F., Kleine, T., Bourdon, B., 2010. Broad bounds on Earth's accretion and core formation constrained by geochemical models. Nature Geoscience. 3, 439-443.
Sanloup, C., Jambon, A., Gillet, P., 1999. A simple chondritic model of Mars. Physics of The Earth and Planetary Interiors. 112, 43-54.
Savage, P., Georg, R., Armytage, R., Williams, H., Halliday, A., 2010. Silicon isotope homogeneity in the mantle. Earth and Planetary Science Letters. 295, 139-146.
Savage, P. S., Armytage, R. M., Georg, R. B., Halliday, A. N., 2014. High temperature silicon isotope geochemistry. Lithos. 190, 500-519.
Savage, P. S., Moynier, F., 2013. Silicon isotopic variation in enstatite meteorites: clues to their origin and Earth-forming material. Earth and Planetary Science Letters. 361, 487-496.
Savage, P. S., Moynier, F., Boyet, M., 2022. Zinc isotope anomalies in primitive meteorites identify the outer solar system as an important source of Earth's volatile inventory. Icarus. 386, 115172.
Schiller, M., Bizzarro, M., Siebert, J., 2020. Iron isotope evidence for very rapid accretion and differentiation of the proto-Earth. Science advances. 6, eaay7604.
Stähler, S. C., et al., 2021. Seismic detection of the martian core. Science. 373, 443-448.
Steele, R. C., Elliott, T., Coath, C. D., Regelous, M., 2011. Confirmation of mass-independent Ni isotopic variability in iron meteorites. Geochimica et Cosmochimica Acta. 75, 7906-7925.
Steller, T., Burkhardt, C., Yang, C., Kleine, T., 2022. Nucleosynthetic zinc isotope anomalies reveal a dual origin of terrestrial volatiles. Icarus. 386, 115171.
Tang, H., Dauphas, N., 2012a. Abundance, distribution, and origin of< sup> 60</sup> Fe in the solar protoplanetary disk. Earth and Planetary Science Letters. 359, 248-263.
Tang, H., Dauphas, N., 2012b. Abundance, distribution, and origin of< sup> 60</sup> Fe in the solar protoplanetary disk. Earth and Planetary Science Letters.
Tang, H., Dauphas, N., 2014. 60Fe-60Ni chronology of core formation on Mars. Earth and Planetary Science Letters. 390, 264-274.
Tissot, F. L., Collinet, M., Namur, O., Grove, T. L., 2022. The case for the angrite parent body as the archetypal first-generation planetesimal: Large, reduced and Mg-enriched. Geochimica et Cosmochimica Acta. 338, 278-301.
Toplis, M. J., et al., 2013. Chondritic models of 4 Vesta: Implications for geochemical and geophysical properties. Meteoritics and Planetary Science. 48, 2300-2315.
Trinquier, A., Birck, J.-L., Allegre, C. J., 2007. Widespread 54Cr heterogeneity in the inner solar system. The Astrophysical Journal. 655, 1179.

Tsiganis, K., Gomes, R., Morbidelli, A., Levison, H. F., 2005. Origin of the orbital architecture of the giant planets of the Solar System. Nature. 435, 459-461.

Walsh, K. J., Morbidelli, A., Raymond, S. N., O'Brien, D. P., Mandell, A. M., 2011. A low mass for Mars from Jupiter/'s early gas-driven migration. Nature. 475, 206-209.

Wänke, H., 1981. Constitution of terrestrial planets. Philosophical Transactions of the Royal Society of London. Series A, Mathematical and Physical Sciences. 303, 287-302.

Ward, W. R., 1986. Density waves in the solar nebula: Diffential Lindblad torque. icarus. 67, 164-180.

Warren, P. H., 2011a. Stable-isotopic anomalies and the accretionary assemblage of the Earth and Mars: A subordinate role for carbonaceous chondrites. Earth and Planetary Science Letters. 311, 93-100.

Warren, P. H., 2011b. Stable-isotopic anomalies and the accretionary assemblage of the Earth and Mars: A subordinate role for carbonaceous chondrites. Earth and Planetary Science Letters. 311, 93-100.

Wasson, J., Kallemeyn, G., 1988. Compositions of chondrites. Philosophical Transactions of the Royal Society of London. Series A, Mathematical and Physical Sciences. 325, 535-544.

Wetherill, G. W., 1991. Why isn't Mars as big as Earth? Lunar and Planetary Science Conference. 22, 1495.

Wombacher, F., Rehkämper, M., Mezger, K., 2004. Determination of the mass-dependence of cadmium isotope fractionation during evaporation. Geochimica et cosmochimica acta. 68, 2349-2357.

Woo, J., Morbidelli, A., Grimm, S., Stadel, J., Brasser, R., 2023. Terrestrial planet formation from a ring. Icarus. 396, 115497.

Young, E. D., Galy, A., Nagahara, H., 2002. Kinetic and equilibrium mass-dependent isotope fractionation laws in nature and their geochemical and cosmochemical significance. Geochimica et Cosmochimica Acta. 66, 1095-1104.

Yu, G., Jacobsen, S. B., 2011. Fast accretion of the Earth with a late Moon-forming giant impact. Proceedings of the National Academy of Sciences. 108, 17604-17609.

Zambardi, T., Poitrasson, F., Corgne, A., Méheut, M., Quitté, G., Anand, M., 2013. Silicon isotope variations in the inner solar system: Implications for planetary formation, differentiation and composition. Geochimica et Cosmochimica Acta. 121, 67-83.

Zhang, J., Huang, S., Davis, A. M., Dauphas, N., Hashimoto, A., Jacobsen, S. B., 2014. Calcium and titanium isotopic fractionations during evaporation. Geochimica et Cosmochimica Acta. 140, 365-380.

Zhu, M.-H., et al., 2021. Common feedstocks of late accretion for the terrestrial planets. Nature Astronomy. 5, 1286-1296.

**Table 1.** Element deliveries in the mantles of Mars and Earth in 3-stage accretion models

| | **Mars (Equilibration=1)** | | Stages | | | **Earth (Equilibration=0.4)** | | Stages | | |
|---|---|---|---|---|---|---|---|---|---|---|
| | | | 0 | 0.6 | 0.992 | | | 0 | 0.6 | 0.995 |
| | $D_{core/mantle}$ | $K_{ic}$ | 0.6 | 0.992 | 1 | $D_{core/mantle}$ | $K_{ic}$ | 0.6 | 0.995 | 1 |
| O | 0 | 0 | 0.600 | 0.392 | 0.008 | 0 | 0 | 0.600 | 0.395 | 0.005 |
| Ca | 0 | 0 | 0.600 | 0.392 | 0.008 | 0 | 0 | 0.600 | 0.395 | 0.005 |
| Ti | 0 | 0 | 0.600 | 0.392 | 0.008 | 0 | 0 | 0.600 | 0.395 | 0.005 |
| Cr | 0 | 0 | 0.600 | 0.392 | 0.008 | 3 | 0.7 | 0.428 | 0.563 | 0.008 |
| Fe | 7 | 2.3 | 0.183 | 0.791 | 0.026 | 14 | 2.6 | 0.157 | 0.825 | 0.018 |
| Ni | 206 | 68.5 | 0.000 | 0.572 | 0.428 | 27 | 5.1 | 0.044 | 0.926 | 0.030 |
| Sr | 0 | 0.0 | 0.600 | 0.392 | 0.008 | 0 | 0.0 | 0.600 | 0.395 | 0.005 |
| Zr | 0 | 0 | 0.600 | 0.392 | 0.008 | 0 | 0 | 0.600 | 0.395 | 0.005 |
| Mo | 14 | 4.7 | 0.055 | 0.900 | 0.044 | 100 | 19.3 | 0.000 | 0.903 | 0.097 |
| Ru | 1129 | 376.3 | 0.000 | 0.000 | 1.000 | 800 | 154.0 | 0.000 | 0.000 | 1.000 |
| Zn | 6 | 2.1 | 0.200 | 0.775 | 0.025 | 0 | 0 | 0.600 | 0.395 | 0.005 |

Note. We assume impactor core-target mantle equilibrations of 1 and 0.4 for Mars and Earth, respectively. The core mass fractions in Mars and Earth are 0.25 and 0.32. Core/mantle partitioning is from McDonough (2003); Brasser et al. (2018); Yoshizaki and McDonough (2020).

**Table 2.** Element deliveries in the mantles of Mars and Earth in 4-stage accretion models

| | Mars (Equilibration=1) | | Stages | | | | Earth (Equilibration=0.4) | | Stages | | | |
|---|---|---|---|---|---|---|---|---|---|---|---|---|
| | | | 0 | 0.5 | 0.75 | 0.992 | | | 0 | 0.5 | 0.75 | 0.995 |
| | $D_{core/mantle}$ | $K_{ic}$ | 0.5 | 0.75 | 0.992 | 1 | $D_{core/mantle}$ | $K_{ic}$ | 0.5 | 0.75 | 0.995 | 1 |
| O | 0 | 0 | 0.500 | 0.250 | 0.242 | 0.008 | 0 | 0 | 0.500 | 0.250 | 0.245 | 0.005 |
| Ca | 0 | 0 | 0.500 | 0.250 | 0.242 | 0.008 | 0 | 0 | 0.500 | 0.250 | 0.245 | 0.005 |
| Ti | 0 | 0 | 0.500 | 0.250 | 0.242 | 0.008 | 0 | 0 | 0.500 | 0.250 | 0.245 | 0.005 |
| Cr | 0 | 0 | 0.500 | 0.250 | 0.242 | 0.008 | 3 | 0.7 | 0.316 | 0.304 | 0.371 | 0.008 |
| Fe | 7 | 2.3 | 0.100 | 0.284 | 0.589 | 0.026 | 14 | 2.6 | 0.081 | 0.271 | 0.630 | 0.018 |
| Ni | 206 | 68.5 | 0.000 | 0.000 | 0.572 | 0.428 | 27 | 5.1 | 0.014 | 0.158 | 0.797 | 0.030 |
| Sr | 0 | 0.0 | 0.500 | 0.250 | 0.242 | 0.008 | 0 | 0.0 | 0.500 | 0.250 | 0.245 | 0.005 |
| Zr | 0 | 0 | 0.500 | 0.250 | 0.242 | 0.008 | 0 | 0 | 0.500 | 0.250 | 0.245 | 0.005 |
| Mo | 14 | 4.7 | 0.020 | 0.176 | 0.760 | 0.044 | 100 | 19.3 | 0.000 | 0.003 | 0.901 | 0.097 |
| Ru | 1129 | 376.3 | 0.000 | 0.000 | 0.000 | 1.000 | 800 | 154.0 | 0.000 | 0.000 | 0.000 | 1.000 |
| Zn | 6 | 2.1 | 0.113 | 0.291 | 0.571 | 0.025 | 0 | 0 | 0.500 | 0.250 | 0.245 | 0.005 |

**Table 3**. Mixing models

| Name | Stages | Endmembers | Isotopic constraints |
|---|---|---|---|
| Baseline model | 3: 0-60, 60-99+%, late veneer | CI, COCV, O, E | O, Ca, Ti, Cr, Fe, Ni, Sr, Zr, Mo, Ru (Earth only) |
| 4-stage model | 4: 0-50, 50-75, 75-99+%, late veneer | CI, COCV, O, E | O, Ca, Ti, Cr, Fe, Ni, Sr, Zr, Mo, Ru (Earth only) |
| Achondrite model | 2: 0-60, 60-100% | CI, COCV, O, E, A | O, Ca, Ti, Cr, Fe, Ni, Sr, Zr, Si |

**Table 4.** Compilation of isotopic anomalies in meteorites and planetary materials

| | $\Delta^{17}O$ | ± | $\varepsilon^{48}Ca$ | ± | $\varepsilon^{50}Ti$ | ± | $\varepsilon^{54}Cr$ | ± | $\mu^{54}Fe$ | ± | $\varepsilon^{64}Ni$ | ± | $\mu^{84}Sr$ | ± | $\mu^{96}Zr$ | ± | $\varepsilon^{92}Mo$ | ± | $\Delta^{95}Mo$ | ± | $\varepsilon^{100}Ru$ | ± | $\mu^{66}Zn$ | ± | $\mu^{30}Si$ | ± |
|---|---|---|---|---|---|---|---|---|---|---|---|---|---|---|---|---|---|---|---|---|---|---|---|---|---|---|
| CI | 0.41 | 0.11 | 2.07 | 0.07 | 1.90 | 0.07 | 1.47 | 0.09 | 0.3 | 1.6 | 0.62 | 0.10 | 42 | 31 | 83 | 16 | 0.97 | 0.45 | 56 | 30 | -0.24 | 0.13 | 32 | 7 | 33.4 | 2.3 |
| CM | -2.92 | 0.44 | 3.13 | 0.14 | 2.93 | 0.11 | 0.98 | 0.06 | 22.8 | 2.7 | 0.37 | 0.09 | 49 | 9 | 104 | 22 | 5.67 | 0.81 | 32 | 15 | -0.71 | 0.91 | 29 | 7 | 16.3 | 2.9 |
| CO | -4.32 | 0.26 | 3.87 | 0.56 | 3.53 | 0.50 | 0.84 | 0.13 | 16.3 | 4.5 | 0.26 | 0.11 | 49 | 15 | 84 | 21 | 2.02 | 0.63 | 15 | 28 | -0.92 | 0.45 | 24 | 2 | 19.0 | 3.0 |
| CV | -3.62 | 0.48 | 2.51 | 0.02 | 3.44 | 0.14 | 0.91 | 0.03 | 26.9 | 3.55 | 0.31 | 0.04 | 71 | 8 | 114 | 13 | 1.80 | 0.07 | 33 | 7 | -1.20 | 0.17 | 29 | 4 | 14.7 | 1.7 |
| CK | -4.46 | 0.21 | na | na | 3.60 | 0.51 | 0.52 | 0.06 | 27.0 | 4.5 | na | na | 38 | 14 | 77 | 36 | 2.27 | 0.42 | 32 | 14 | -1.10 | 0.16 | na | na | 21.3 | 3.9 |
| CR | -1.48 | 0.55 | 2.15 | 0.11 | 2.41 | 0.25 | 1.28 | 0.06 | 32.4 | 4.6 | 0.26 | 0.15 | -1 | 5 | 119 | 32 | 3.02 | 0.81 | 16 | 30 | -0.86 | 0.20 | 30 | 3 | 9.8 | 3.4 |
| CH | -1.55 | 0.27 | na | na | na | na | 1.49 | 0.05 | 15.5 | 6.6 | na | na | -4 | 25 | 111 | 9 | 1.88 | 0.54 | 23 | 4 | -0.91 | 0.13 | na | na | | |
| CB | -2.14 | 0.38 | na | na | 2.04 | 0.07 | 1.28 | 0.05 | na | na | na | na | na | na | 114 | 15 | 1.74 | 0.25 | 17 | 18 | 1.05 | 0.09 | na | na | | |
| CL | -4.31 | 0.57 | na | na | 2.59 | 0.14 | 0.74 | 0.06 | na | na | na | na | na | na | 89 | 18 | na | na | na | na | na | na | na | na | | |
| H | 0.72 | 0.05 | -0.15 | 0.26 | -0.63 | 0.13 | -0.38 | 0.02 | 11.8 | 1.9 | -0.16 | 0.05 | 2 | 38 | 44 | 14 | 0.86 | 0.19 | -14 | 6 | -0.28 | 0.04 | -19 | 9 | | |
| L | 1.03 | 0.04 | -0.14 | 0.02 | -0.63 | 0.02 | -0.39 | 0.07 | 10.6 | 4.7 | -0.11 | 0.06 | 4 | 14 | 34 | 13 | 0.69 | 0.24 | -15 | 14 | -0.28 | 0.13 | -21 | 5 | 7.4 | 4.3 |
| LL | 1.19 | 0.06 | -0.36 | 0.05 | -0.66 | 0.04 | -0.42 | 0.05 | 12.9 | 4.0 | -0.18 | 0.05 | -15 | 15 | 30 | 4 | 0.77 | 0.30 | -16 | 8 | 0.01 | 0.12 | -18 | 6 | 7.4 | 2.9 |
| R | 2.64 | 0.15 | 0.08 | 0.04 | na | na | -0.03 | 0.07 | 9.3 | 3.6 | -0.08 | 0.10 | -16 | 6 | 26 | 3 | 0.77 | 0.35 | -9 | 6 | -0.39 | 0.13 | na | na | 1.1 | 3.1 |
| EH | -0.03 | 0.10 | -0.11 | 0.02 | -0.11 | 0.03 | 0.04 | 0.05 | 5.7 | 2.0 | 0.14 | 0.11 | -14 | 13 | 15 | 6 | 0.48 | 0.33 | -11 | 9 | -0.08 | 0.07 | -16 | 3 | 11.8 | 2.7 |
| EL | -0.01 | 0.07 | -0.18 | 0.04 | -0.28 | 0.05 | 0.04 | 0.05 | 6.7 | 1.4 | -0.05 | 0.07 | 6 | 25 | 17 | 7 | 0.37 | 0.09 | -7 | 5 | -0.08 | 0.05 | -17 | 4 | 11.8 | 3.1 |
| K | -1.09 | 0.35 | -1.30 | 0.25 | na | na | -0.44 | 0.04 | na | na | na | na | na | na | 42 | 7 | na | na | na | na | na | na | na | na | | |
| Acapulcoite | -0.99 | 0.09 | na | na | -1.31 | 0.05 | -0.63 | 0,17 | na | na | na | na | na | na | 32 | 4 | 1.04 | 0.11 | -8 | 7 | -0.33 | 0.05 | na | na | | |
| Lodranite | -1.11 | 0.11 | na | na | -1.68 | 0.31 | na | na | na | na | na | na | na | na | na | na | 1.06 | 0.13 | -5 | 6 | -0.37 | 0.11 | na | na | | |
| Brachinite | -0.26 | 0.08 | na | na | -1.38 | 0.13 | -0.45 | 0.06 | na | na | na | na | na | na | na | na | 1.34 | 0.10 | -12 | 7 | 0.08 | 0.18 | na | na | | |
| Winonaite | -0.49 | 0.07 | -0.21 | 0.09 | na | na | na | na | na | na | na | na | na | na | 2 | 9 | 0.35 | 0.79 | -5 | 12 | -0.06 | 0.10 | na | na | | |
| NWA5363/5400 | -0.01 | 0.02 | -0.53 | 0.20 | -1.02 | 0.10 | -0.37 | 0.13 | na | na | 0.04 | 0.08 | na | na | na | na | 0.53 | 0.22 | -30 | 19 | -0.34 | 0.13 | na | na | | |
| Angrite | -0.11 | 0.11 | -1.04 | 0.18 | -1.15 | 0.02 | -0.43 | 0.05 | na | na | 0.09 | 0.07 | -2 | 9 | 50 | 5 | 0.47 | 0.57 | -6 | 9 | na | na | na | na | -10.1 | 2.0 |
| Aubrite | 0.06 | 0.14 | -0.44 | 0.59 | -0.05 | 0.04 | 0.05 | 0.05 | na | na | 0.14 | 0.33 | na | na | 21 | 4 | 0.56 | 0.13 | -4 | 6 | -0.06 | 0.03 | na | na | | |
| HED | -0.29 | 0.08 | -1.21 | 0.23 | -1.23 | 0.04 | -0.67 | 0.07 | 12.0 | 2.5 | 0.18 | 0.10 | -7 | 8 | 39 | 8 | na | na | na | na | na | na | na | na | -9.2 | 3.0 |
| Ureilite | -1.16 | 0.15 | -1.36 | 0.03 | -1.97 | 0.14 | -0.88 | 0.03 | 15.5 | 1.5 | -0.20 | 0.21 | na | na | na | na | 1.08 | 0.16 | -14 | 4 | -0.30 | 0.05 | -34 | 2 | -9.5 | 2.8 |
| Moon | 0.004 | 0.003 | 0.04 | 0.02 | -0.03 | 0.04 | 0.09 | 0.08 | 2.0 | 2.0 | na | na | 17 | 32 | 8 | 8 | na | na | | | na | na | na | na | | |
| Mesosiderite | -0.24 | 0.06 | na | na | -1.26 | 0.10 | -0.70 | 0.06 | na | na | na | na | na | na | 35 | 6 | 1.28 | 0.16 | -16 | 7 | -0.42 | 0.02 | na | na | -8.0 | 3.7 |
| MG pal | -0.28 | 0.03 | na | na | -1.37 | 0.08 | -0.39 | 0.46 | na | na | -0.13 | 0.19 | na | na | 43 | 18 | 1.12 | 0.05 | -14 | 2 | -0.47 | 0.11 | na | na | -8.1 | 1.7 |
| ESPAL | -4.69 | 0.10 | na | na | na | na | 0.71 | 0.18 | na | na | na | na | na | na | na | na | 1.49 | 0.33 | 32 | 34 | na | na | na | na | | |
| IAB Complex (IAB+IIICD) | -0.48 | 0.05 | na | na | na | na | na | na | -1.0 | 3.9 | -0.07 | 0.06 | na | na | na | na | 0.19 | 0.17 | -11 | 3 | -0.03 | 0.01 | -19 | 13 | | |
| IC | na | na | na | na | na | na | na | na | 12.9 | 4.6 | -0.24 | 0.09 | na | na | na | na | 0.96 | 0.06 | -21 | 2 | -0.28 | 0.03 | -30 | 3 | | |
| IIAB | -0.66 | 0.13 | na | na | na | na | -0.82 | 0.08 | 20.6 | 5.5 | -0.11 | 0.27 | na | na | na | na | 1.41 | 0.12 | -20 | 3 | -0.43 | 0.02 | na | na | | |
| IIC | na | na | na | na | na | na | na | na | 35.1 | 3.7 | 0.39 | 0.10 | na | na | na | na | 3.08 | 0.17 | 22 | 4 | -0.28 | 0.03 | na | na | | |
| IID | na | na | na | na | na | na | na | na | 24.4 | 4.9 | 0.20 | 0.07 | na | na | na | na | 1.54 | 0.16 | 28 | 6 | -1 | 0.05 | 23 | 3 | | |
| IIE | 0.57 | 0.05 | na | na | na | na | -0.59 | 0.13 | na | na | na | na | na | na | na | na | 0.91 | 0.24 | -13 | 8 | na | na | na | na | | |
| IIF | na | na | na | na | na | na | na | na | na | na | 0.17 | 0.11 | na | na | na | na | 1.52 | 0.15 | 31 | 6 | -1 | 0.06 | na | na | | |
| IIG | na | na | na | na | na | na | na | na | na | na | na | na | na | na | na | na | na | na | na | na | na | na | na | na | | |
| IIIAB | -0.20 | 0.04 | na | na | na | na | -0.81 | 0.05 | 9.8 | 3.5 | -0.27 | 0.15 | na | na | na | na | 1.18 | 0.06 | -19 | 2 | -0.6 | 0.02 | -33 | 5 | | |
| IIIE | na | na | na | na | na | na | na | na | na | na | -0.26 | 0.08 | na | na | na | na | 1.10 | 0.11 | -15 | 5 | -0.52 | 0.06 | na | na | | |
| IIIF | na | na | na | na | na | na | na | na | 27.0 | 7.0 | 0.27 | 0.09 | na | na | na | na | 1.53 | 0.17 | 22 | 5 | -0.91 | 0.11 | na | na | | |
| IVA | 1.17 | 0.10 | na | na | na | na | -0.47 | 0.06 | 9.0 | 3.9 | -0.27 | 0.07 | na | na | na | na | 0.99 | 0.07 | -12 | 5 | -0.32 | 0.04 | na | na | | |
| IVB | na | na | na | na | na | na | na | na | na | na | 0.21 | 0.04 | na | na | na | na | 1.97 | 0.07 | 16 | 3 | -0.92 | 0.02 | na | na | | |
| Silicate Mars | 0.27 | 0.03 | -0.20 | 0.02 | -0.44 | 0.06 | -0.15 | 0.03 | 6.6 | 2.2 | 0.14 | 0.10 | -27 | 12 | 27 | 2 | 0.57 | 0.08 | 5 | 5 | na | na | -21 | 3 | -5.5 | 1.9 |
| Silicate Earth | 0.002 | 0.004 | 0.01 | 0.01 | 0.01 | 0.01 | 0.09 | 0.12 | 0.0 | 1.0 | 0.12 | 0.02 | -0.5 | 6.6 | 1 | 2 | 0.08 | 0.13 | 8 | 5 | 0.02 | 0.01 | -4.3 | 3.0 | 1.3 | 2.5 |
| **Endmember and calculated target compositions** | | | | | | | | | | | | | | | | | | | | | | | | | | |
| CI | 0.41 | 0.11 | 2.07 | 0.07 | 1.90 | 0.07 | 1.47 | 0.09 | 0.3 | 1.6 | 0.62 | 0.10 | 42 | 31 | 83 | 16 | 0.97 | 0.45 | 56 | 30 | -0.24 | 0.13 | 32 | 7 | 33.4 | 2.3 |
| COCV | -3.90 | 0.33 | 2.51 | 0.02 | 3.45 | 0.14 | 0.89 | 0.04 | 21.8 | 4.5 | 0.31 | 0.03 | 66 | 7 | 103 | 13 | 1.9 | 0.06 | 26 | 4 | -1.16 | 0.17 | 25 | 1 | 15.7 | 1.5 |
| O | 0.99 | 0.04 | -0.17 | 0.02 | -0.64 | 0.02 | -0.39 | 0.02 | 12.1 | 1.7 | -0.16 | 0.03 | -2 | 13 | 37 | 7 | 0.81 | 0.13 | -15 | 4 | -0.22 | 0.07 | -19 | 4 | 7.4 | 2.4 |
| E | -0.02 | 0.07 | -0.12 | 0.02 | -0.16 | 0.06 | 0.04 | 0.04 | 6.4 | 1.1 | 0.06 | 0.09 | -11 | 13 | 14 | 5 | 0.40 | 0.08 | -8 | 4 | -0.08 | 0.04 | -16 | 2 | 12.5 | 2.2 |
| A (HED) | -0.29 | 0.08 | -1.21 | 0.23 | -1.23 | 0.04 | -0.67 | 0.07 | 12.0 | 2.5 | 0.18 | 0.10 | -7 | 8 | 39 | 8 | na | na | na | na | na | na | na | na | -9.2 | 3.0 |
| CI+COCV+O+E mixing | | | | | | | | | | | | | | | | | | | | | | | | | | |
| Baseline Mars | 0.28 | 0.03 | -0.20 | 0.02 | -0.32 | 0.04 | -0.08 | 0.04 | 7.2 | 1.0 | 0.13 | 0.08 | -14 | 7 | 27 | 2 | 0.55 | 0.06 | -2 | 4 | -0.45 | 0.38 | | | | |
| 4-stage Mars | 0.28 | 0.03 | -0.20 | 0.02 | -0.31 | 0.04 | -0.08 | 0.04 | 7.4 | 1.0 | 0.13 | 0.08 | -13 | 7 | 27 | 2 | 0.56 | 0.06 | -2 | 4 | -0.47 | 0.40 | | | | |
| Baseline Earth | 0.002 | 0.004 | 0.01 | 0.03 | 0.01 | 0.01 | 0.13 | 0.05 | 3.0 | 0.7 | 0.12 | 0.02 | -2 | 6 | 4 | 2 | 0.34 | 0.08 | 7 | 4 | 0.02 | 0.01 | | | | |
| 4-stage Earth | 0.002 | 0.004 | 0.01 | 0.02 | 0.01 | 0.01 | 0.14 | 0.05 | 3.0 | 0.7 | 0.12 | 0.02 | -2 | 6 | 4 | 2 | 0.34 | 0.08 | 6 | 5 | 0.01 | 0.01 | | | | |
| CI+COCV+O+E+A mixing | | | | | | | | | | | | | | | | | | | | | | | | | | |
| Achondrite Mars | 0.26 | 0.03 | -0.20 | 0.02 | -0.48 | 0.05 | -0.25 | 0.08 | 8.8 | 1.5 | 0.17 | 0.09 | -5 | 6 | 29 | 2 | | | | | | | | | -0.08 | 1.63 |
| Achondrite Earth | 0.002 | 0.004 | 0.01 | 0.01 | 0.01 | 0.01 | 0.11 | 0.04 | 3.0 | 0.7 | 0.12 | 0.02 | -2 | 6 | 4 | 2 | | | | | | | | | 8.134 | 1.549 |

**Table 5.** Concentrations (in ppm) in major chondrite groups (compiled from Lodders 2021)

| | O | Ca | Ti | Cr | Fe | Ni | Sr | Zr | Mo | Ru | Zn | Si |
|---|---|---|---|---|---|---|---|---|---|---|---|---|
| CI | 453840 | 8840 | 450 | 2610 | 185620 | 10950 | 7.79 | 3.79 | 0.976 | 0.666 | 311 | 107740 |
| CM | 424100 | 12000 | 610 | 3090 | 211000 | 12500 | 10 | 5.2 | 1.2 | 0.86 | 177 | 132000 |
| CO | 364200 | 15700 | 750 | 3550 | 248000 | 14100 | 13.2 | 6 | 1.6 | 1.03 | 101 | 158000 |
| CV | 367800 | 18000 | 850 | 3600 | 236000 | 14000 | 14.6 | 6.8 | 1.45 | 1.07 | 113 | 159000 |
| CR | 392700 | 13400 | 600 | 3640 | 237700 | 13600 | 10 | 5.2 | 1.3 | 0.91 | 65 | 150000 |
| H | 337400 | 12400 | 630 | 3500 | 271500 | 17100 | 8.7 | 5.7 | 1.4 | 1.1 | 47 | 170400 |
| L | 368400 | 13000 | 690 | 3660 | 218450 | 12700 | 11 | 5.8 | 1.2 | 0.75 | 54 | 184600 |
| LL | 383700 | 13200 | 680 | 3680 | 198000 | 10200 | 11 | 6.8 | 1.1 | 0.55 | 55 | 189000 |
| EH | 308000 | 8500 | 480 | 3140 | 311000 | 18700 | 7.1 | 3.8 | 1 | 0.91 | 320 | 167300 |
| EL | 332100 | 10000 | 510 | 3200 | 250000 | 15400 | 7.9 | 4.8 | 0.73 | 0.92 | 16 | 190000 |

**Table 6.** Iron isotopic compositions of Martian meteorites

| Sample | N[a] | $\mu^{54}Fe_{(7/6)}$[b] | $\mu^{58}Fe_{(7/6)}$ | $\delta^{56}Fe$[c] |
|---|---|---|---|---|
| *This study* | | | | |
| Chassigny | 13 | 8±7 | -7±9 | 0.00±0.02 |
| Nakhla | 15 | 4±5 | -5±6 | 0.02±0.02 |
| Shergotty | 12 | 7±6 | -4±7 | 0.03±0.02 |
| Zagami | 15 | 6±5 | -3±12 | 0.03±0.02 |
| *Schiller et al. (2020)* | | | | |
| Zagami | | 6±6 | | -0.05±0.02 |
| Replicate | | 8±7 | | -0.02±0.03 |
| Tissinit | | 5±7 | | 0.05±0.04 |
| Replicate | | 7±6 | | 0.01±0.01 |
| Bulk Silicate Mars[d] | | 6±2 | -5±3 | 0.00±0.03 |

[a] Number of individual analyses.

[b] Fe isotopic composition internally normalized to $^{57}Fe/^{56}Fe$=0.023095 and expressed as μ-notation defined as the parts-per-million deviation of the $^{5x}Fe/^{56}Fe$ ratio in the sample relative to the two IRMM-524a standard solution bracketing measurements.

[c] Mass-dependent Fe isotopic composition calculated by sample-standard bracketing and given as δ-notation defined parts-per-thousands deviation of the $^{56}Fe/^{54}Fe$ ratio of the sample relative to the IRMM-524a standard solution.

[d] Weighted average Fe isotopic composition of bulk silicate Mars calculated using data from this study and Schiller et al. (2020).

**Table 7**. Building blocks of Mars (contributions in wt%)

| Stage | Span (%) | CI | COCV | O | E | O+E=NC |
|---|---|---|---|---|---|---|
| I | 0<br>60 | 0.7 +2.0 -0.5 | 0.4 +1.0 -0.3 | 47.0 +12.8 -13.0 | 51.9 +13.2 -13.0 | 99.0 +0.7 -2.0 |
| II | 60<br>99.2 | 2.8 +5.1 -2.3 | 0.7 +1.7 -0.6 | 11.0 +15.8 -9.5 | 85.4 +10.6 -17.1 | 96.5 +2.6 -4.8 |
| III-late veneer | 99.2<br>100 | 32.5 +41.3 -28.5 | 23.4 +49.4 -18.2 | 20.9 +40.6 -17.0 | 23.3 +47.4 -18.3 | 44.1 +39.0 -31.2 |
| Bulk | 0<br>100 | 1.8 +2.0 -1.2 | 0.7 +0.7 -0.5 | 32.7 +5.3 -4.9 | 64.8 +5.1 -5.4 | 97.5 +1.2 -1.9 |

| Stage | Span (%) | CI | COCV | O | E | O+E=NC |
|---|---|---|---|---|---|---|
| I | 0<br>50 | 0.7 +2.1 -0.5 | 0.4 +1.0 -0.3 | 45.0 +20.9 -23.3 | 53.9 +23.5 -20.9 | 98.9 +0.8 -2.1 |
| II | 50<br>75 | 1.7 +5.1 -1.1 | 0.8 +2.2 -0.5 | 32.3 +50.1 -28.1 | 65.2 +28.3 -50.5 | 97.5 +1.8 -5.1 |
| III | 75<br>99.2 | 4.6 +6.8 -4.0 | 1.0 +2.5 -0.7 | 10.2 +18.3 -8.4 | 84.2 +11.0 -19.6 | 94.4 +4.3 -6.6 |
| IV-late veneer | 99.2<br>100 | 26.4 +36.8 -23.2 | 25.9 +52.3 -20.0 | 22.8 +41.8 -18.7 | 24.9 +44.5 -20.7 | 47.7 +37.2 -35.0 |
| Bulk | 0<br>100 | 1.9 +1.8 -1.3 | 0.6 +0.7 -0.4 | 33.1 +5.4 -5.4 | 63.6 +5.5 -5.5 | 96.7 +1.3 -1.7 |

| Stage | Span (%) | CI | COCV | O | E | A | O+E+A=NC |
|---|---|---|---|---|---|---|---|
| I | 0<br>60 | 5.5 +13.7 -4.0 | 2.2 +5.6 -1.5 | 52.3 +32.3 -19.2 | 4.8 +15.6 -3.3 | 35.1 +21.2 -32.5 | 92.3 +5.8 -11.7 |
| II | 60<br>100 | 21.5 +15.1 -21.4 | 2.1 +6.4 -1.4 | 15.6 +24.6 -13.2 | 17.0 +22.5 -15.2 | 43.8 +27.2 -33.7 | 76.3 +20.7 -14.4 |
| Bulk | 0<br>100.0 | 11.9 +8.0 -10.4 | 2.2 +3.3 -1.7 | 37.6 +16.4 -11.4 | 9.7 +11.2 -7.5 | 38.6 +8.2 -12.8 | 85.9 +7.8 -6.7 |

**Table 8**. Building blocks of Earth (contributions in wt%)

| Stage | Span (%) | CI | COCV | O | E | O+E=NC |
|---|---|---|---|---|---|---|
| I | 0<br>60 | 2.1 (+4.5 / -1.6) | 0.7 (+1.3 / -0.6) | 1.0 (+3.0 / -0.6) | 96.2 (+2.6 / -5.2) | 97.2 (+2.1 / -4.4) |
| II | 60<br>99.5 | 11.2 (+6.5 / -6.8) | 0.3 (+0.8 / -0.2) | 0.6 (+1.9 / -0.4) | 87.9 (+6.9 / -6.8) | 88.5 (+6.8 / -6.5) |
| III-late veneer | 99.5<br>100 | 90.4 (+6.9 / -14.8) | 0.6 (+1.7 / -0.4) | 2.8 (+7.4 / -2.0) | 6.2 (+15.0 / -4.5) | 9.0 (+14.9 / -6.9) |
| Bulk | 0<br>100 | 6.1 (+2.4 / -2.4) | 0.6 (+0.8 / -0.4) | 0.8 (+1.9 / -0.6) | 92.5 (+2.3 / -2.5) | 93.3 (+2.1 / -2.2) |

| Stage | Span (%) | CI | COCV | O | E | O+E=NC |
|---|---|---|---|---|---|---|
| I | 0<br>50 | 1.8 (+4.3 / -1.3) | 0.7 (+1.5 / -0.5) | 1.2 (+3.5 / -0.8) | 96.3 (+2.5 / -4.8) | 97.5 (+1.8 / -4.3) |
| II | 50<br>75 | 3.6 (+10.0 / -2.4) | 0.9 (+2.6 / -0.6) | 2.2 (+9.4 / -1.2) | 93.3 (+4.4 / -11.7) | 95.5 (+3.0 / -10.0) |
| III | 75<br>99.5 | 19.1 (+9.9 / -10.1) | 0.3 (+0.9 / -0.2) | 0.7 (+2.1 / -0.5) | 79.9 (+10.1 / -10.0) | 80.6 (+10.0 / -9.8) |
| IV-late veneer | 99.5<br>100 | 2.5 (+8.3 / -1.6) | 0.3 (+0.8 / -0.2) | 1.5 (+4.4 / -1.0) | 95.7 (+2.9 / -8.7) | 97.2 (+1.7 / -8.2) |
| Bulk | 0<br>100 | 6.5 (+2.7 / -2.6) | 0.6 (+0.8 / -0.5) | 1.3 (+2.7 / -0.9) | 91.0 (+2.7 / -4.0) | 92.4 (+2.3 / -2.6) |

| Stage | Span (%) | CI | COCV | O | E | A | O+E+A=NC |
|---|---|---|---|---|---|---|---|
| I | 0<br>60 | 2.5 (+3.1 / -2.5) | 1.2 (+1.7 / -1.0) | 0.9 (+2.4 / -0.6) | 94.7 (+3.5 / -3.5) | 0.7 (+1.9 / -0.4) | 96.3 (+2.9 / -2.9) |
| II | 60<br>100 | 7.9 (+6.1 / -5.2) | 0.6 (+1.6 / -0.4) | 0.7 (+2.0 / -0.5) | 90.2 (+5.6 / -6.1) | 0.6 (+1.6 / -0.4) | 91.5 (+5.2 / -5.9) |
| Bulk | 0<br>100.0 | 4.7 (+1.6 / -2.4) | 0.9 (+0.9 / -0.7) | 0.8 (+1.6 / -0.6) | 92.9 (+2.1 / -2.3) | 0.6 (+1.2 / -0.5) | 94.4 (+1.6 / -1.2) |

**Table 9.** Comparison between prior and posterior probabilities for the nuisance parameters (the isotopic compositions of the endmembers)

| | $\Delta^{17}$O | ± | $\varepsilon^{48}$Ca | ± | $\varepsilon^{50}$Ti | ± | $\varepsilon^{54}$Cr | ± | $\varepsilon^{54}$Fe | ± | $\varepsilon^{64}$Ni | ± | $\mu^{84}$Sr | ± | $\varepsilon^{96}$Zr | ± | $\varepsilon^{92}$Mo | ± | $\Delta^{95}$Mo | ± | $\varepsilon^{100}$Ru | ± |
|---|---|---|---|---|---|---|---|---|---|---|---|---|---|---|---|---|---|---|---|---|---|---|
| Prior for the chondritic endmembers | | | | | | | | | | | | | | | | | | | | | | |
| CI | 0.41 | 0.11 | 2.07 | 0.07 | 1.90 | 0.07 | 1.47 | 0.09 | 0.3 | 1.6 | 0.62 | 0.10 | 42 | 31 | 83 | 16 | 0.97 | 0.45 | 56 | 30 | -0.24 | 0.13 |
| COCV | -3.90 | 0.33 | 3.62 | 0.48 | 3.45 | 0.14 | 0.89 | 0.04 | 21.8 | 4.5 | 0.31 | 0.03 | 66 | 7 | 103 | 13 | 1.90 | 0.06 | 26 | 4 | -1.16 | 0.17 |
| O | 0.99 | 0.04 | -0.32 | 0.08 | -0.64 | 0.02 | -0.39 | 0.02 | 12.1 | 1.7 | -0.16 | 0.03 | -2 | 13 | 37 | 7 | 0.81 | 0.13 | -15 | 4 | -0.22 | 0.07 |
| E | -0.02 | 0.07 | -0.25 | 0.15 | -0.16 | 0.06 | 0.04 | 0.04 | 6.4 | 1.1 | 0.06 | 0.09 | -11 | 13 | 14 | 5 | 0.40 | 0.08 | -8 | 4 | -0.08 | 0.04 |
| Posterior in Mars baseline model | | | | | | | | | | | | | | | | | | | | | | |
| CI | 0.41 | 0.11 | 2.07 | 0.07 | 1.90 | 0.07 | 1.47 | 0.09 | 0.3 | 1.6 | 0.62 | 0.10 | 41 | 31 | 83 | 16 | 1.00 | 0.44 | 66 | 29 | -0.24 | 0.13 |
| COCV | -3.91 | 0.33 | 3.61 | 0.48 | 3.44 | 0.14 | 0.89 | 0.04 | 21.6 | 4.4 | 0.31 | 0.03 | 66 | 6 | 103 | 13 | 1.90 | 0.06 | 26 | 4 | -1.16 | 0.17 |
| O | 0.98 | 0.04 | -0.33 | 0.08 | -0.65 | 0.02 | -0.39 | 0.01 | 12.0 | 1.7 | -0.16 | 0.03 | -8 | 12 | 39 | 6 | 0.82 | 0.13 | -14 | 4 | -0.22 | 0.06 |
| E | -0.07 | 0.06 | -0.25 | 0.08 | **-0.23** | **0.05** | 0.04 | 0.04 | 6.3 | 1.0 | 0.06 | 0.08 | **-20** | **11** | 16 | 4 | 0.42 | 0.07 | -4 | 4 | -0.08 | 0.04 |
| Posterior in Earth baseline model | | | | | | | | | | | | | | | | | | | | | | |
| CI | 0.41 | 0.10 | 2.07 | 0.07 | 1.90 | 0.07 | 1.46 | 0.09 | -0.3 | 1.6 | 0.62 | 0.09 | 44 | 28 | 74 | 15 | 0.36 | 0.41 | 64 | 22 | **0.05** | **0.03** |
| COCV | -3.88 | 0.31 | 3.60 | 0.47 | 3.45 | 0.12 | 0.89 | 0.04 | 21.4 | 4.3 | 0.31 | 0.03 | 66 | 7 | 102 | 13 | 1.90 | 0.06 | 26 | 4 | -1.16 | 0.17 |
| O | 0.99 | 0.04 | -0.32 | 0.09 | -0.64 | 0.02 | -0.39 | 0.01 | 12.0 | 1.7 | -0.16 | 0.03 | -2 | 13 | 37 | 7 | 0.81 | 0.13 | -15 | 4 | -0.21 | 0.06 |
| E | -0.02 | 0.04 | -0.16 | 0.06 | -0.13 | 0.05 | 0.04 | 0.04 | **3.2** | **0.8** | 0.07 | 0.03 | -6 | 6 | **-2** | **2** | **0.32** | **0.07** | -7 | 4 | -0.08 | 0.04 |

Note. The values highlighted in bold font correspond to values that change by more than 1$\sigma$ in the posterior relative to the prior.